\documentclass[amsmath,trackchanges,twocolumn]{aastex702}
\usepackage[T1]{fontenc}

\shorttitle{Binary pulsars follow-up}
\shortauthors{Mignani et al.}

\begin{document}

\newcommand{\sw}{{\em Swift}}
\newcommand{\xmm}{{\em XMM-Newton}} 
\newcommand{\gal}{{\em GALEX}}

\def\Pspin{P$_{\rm s}$}
\def\Pdot{\dot {P}}
\def\lapp{\ifmmode\stackrel{<}{_{\sim}}\else$\stackrel{<}{_{\sim}}$\fi}


 \shorttitle{Binary pulsar follow ups}
\title[Binary pulsar follow ups]{Binary pulsar studies with multi-wavelength sky surveys. \\ II. Companion star follow ups\footnote{At the time of writing, RPM is hospitalised and the submission of this manuscript is due after his release from the hospital} }

\author[0000-0002-8685-583X]{R. P. Mignani}
\affiliation{INAF Istituto di Astrofisica Spaziale e Fisica Cosmica Milano, via A. Corti 12, 20133, Milano, Italy}
\affiliation{Janusz Gil Institute of Astronomy, University of Zielona G\'ora, ul Szafrana 2, 65-265, Zielona G\'ora, Poland}
\email[show]{roberto.mignani@inaf.it} 

\author[0000-0002-7104-2107]{C. Pallanca}
\affiliation{Dipartimento di Fisica e Astronomia, Universit\`a degli Studi di Bologna, Viale Berti Pichat 6-2, 40127, Bologna, Italy}
\email{cristina.pallanca3@unibo.it}

\author[0000-0003-1033-1340 ]{V. Testa}
\affiliation{INAF Osservatorio Astronomico di Roma, Via Frascati 33, 00078 Monte Porzio Catone, Roma, Italy}
\email{vincenzo.testa@inaf.it}

\author[0000-0002-5924-3141]{A. Corongiu}
\affiliation{INAF Osservatorio Astronomico di Cagliari,  Via della Scienza 5, 09047, Selargius, Cagliari, Italy}
\email{alessandro.corongiu@inaf.it}

\begin{abstract}
The identification of the stellar companions to binary pulsars and the extraction of the physical information encoded in their optical spectra are key to study the evolution of the binary system and how this is influenced by the interactions between the two stars. Obviously, extending the characterisation of the companion star spectra over an as large as possible wavelength range allows one to maximise the amount of information.  In Paper I, we used 11 source catalogues built upon multi-wavelength (optical, infrared, ultraviolet) imaging sky surveys to search for the stellar companions of 144 field binary pulsars (i.e. not in Globular Clusters) selected from the Australia Telescope National Facility (ATNF) pulsar catalogue.
For nine of these pulsars, for which the companion star had been already identified,  we obtained new spectral information  from imaging photometry in  the surveys' bands. In particular, for some of these pulsars we provided the first possible detections of the companion star in either the ultraviolet or mid-infrared. Here, we serve this additional flux information as a base to update the spectral energy distribution (SED) of the companion stars and the multi-epoch catalogue photometry to look for flux modulation at the orbital period of the binary system.
\end{abstract}


\section{Introduction}

The study of the optical spectra of the companion stars to binary radio pulsars, either via spectroscopy or broad-band photometry, is key to determine their classification, their evolutionary stage, and their physical parameters, such as surface temperature, surface gravity, and chemical composition of the atmosphere. This information is important to reconstruct the evolutionary history of the binary system, independently determine the pulsar age, and study  the interaction between the two stars, including the ablation process of the companion through irradiation from the pulsar relativistic wind, which occurs in the so-called Black Widow and Red Backs systems (see, e.g. Roberts 2013 for a summary).   In particular,  these observations are important to study binary systems where the pulsar undergoes episodes of matter accretion from the companion, occasionally turning into a radio-quiet, but X-ray active, Low-Mass X-ray Binaries (LMXBs). The characterisation of the spectrum of the companion star over a  wavelength range as wide as possible allows one to maximise the amount of  information, hence the scientific return.

\begin{table*}
\begin{center}
\caption{Summary of the properties of the imaging detectors employed in the different surveys that produced the source catalogues used in Paper I. Columns list the detector field--of--view (FoV), pixel scale, the filters peak wavelengths in $\mu$m, and the nominal $1 \sigma$ accuracy $\delta r$ on the catalogue absolute astrometry.  Surveys are listed in order of increasing wavelength.}
\label{survey}
\begin{tabular}{lcclc} \hline
Survey  & FoV & Pixel Scale & Filter central wavelengths ($\mu$m)   & $\delta r$ \\ \hline
{\em GALEX}      & 1\fdg2$\times$1\fdg2     &    1\farcs5       & NUV (0.27), FUV (0.15) & 0\farcs5\\
{\em EUVE}	 & r=4--5$^{\circ}$                  &  10\farcs71--8\farcs48  &  Lexan/Boron (0.011), Al/Ti/C (0.019), Ti/Sb/Al (0.047), Sn/SiO (0.062) & 15\arcsec\\               
SDSS        & 2\fdg5$\times$2\fdg           & 0\farcs396 & u (0.35), g (0.47), r (0.62), i (0.76), z (0.91) & 0\farcs1\\ 
{\em XMM}/OM & 17\arcmin$\times$17\arcmin & 0\farcs5 & U (0.34), B (0.45), V (0.54), UVW1 (0.29), UVM2 (0.23), UVW2 (0.21) & 0\farcs7\\
{\em Swift}/UVOT     &  17\arcmin$\times17$\arcmin & 0\farcs5 & U (0.34), B (0.44), V (0.54), UVW1 (0.26), UVM2 (0.22), UVW2 (0.19)  & 0\farcs5 \\
Catalina$^*$	      &2\fdg8$\times$2\fdg8               &2\farcs5 & V (0.55) & 0\farcs5 \\
                       & 1\fdg2$\times1$\fdg2            & 1\farcs0 & V (0.55) & \\ 
                       & 2\fdg0$\times2$\fdg0          & 1\farcs8 & V (0.55) &  \\
UKIDSS	     & 0\fdg22$\times$0\fdg22             & 0\farcs4 & Z (0.88), Y (1.03), J (1.25), H (1.63), K (2.20) & 0\farcs1 \\
VISTA	     & 1\fdg5$\times1^{\circ}$             &  0\farcs34 & Z (0.88), Y (1.02), J (1.25), H (1.65), K$_{\rm s}$ (2.15) & 0\farcs3\\ 
{\em Spitzer}$^\dagger$	     & 5\arcmin$\times$5\arcmin        & 1\farcs2   & 3.6, 4.5, 5.8, 8.0  & 0\farcs25\\
{\em WISE} 	     &  47\arcmin$\times$47\arcmin\  & 2\farcs75 &  W1 (3.4), W2 (4.6), W3 (12), W4 (22) & 0\farcs5 \\
{\em AKARI}$^\ddagger$ 	     & 9\farcm5$\times$10\arcmin &1\farcs46   &   N2 (2.4), N3 (3.2), N4 (4.1) & 0\farcs4\\
	  	        & 9\farcm1$\times$10\arcmin  & 2\farcs34 &  S7 (7.0), S9W (9.0), S11 (11.0) & \\
                        & 10\farcm3$\times$10\farcm2 & 2\farcs51$\times$2\farcs39  &  L15 (15.0), L18W (18.0) L24 (24.0) & \\
 \hline	 
\end{tabular}
\end{center}
$^*$The survey includes data from the 0.7m Schmidt (Mt. Bigelow, Arizona), the 1.5m  (Mt. Lemmon, Arizona), and the 0.5m Schmidt  (Siding Spring, Australia) telescopes (Drake et al.\ 2009), top to bottom, respectively. \\
$^\dagger$The survey only includes data from the Infrared Array Camera (IRAC; Fazio et al.\ 2004).\\
$^\ddagger$ The survey includes data from all the three channels (NIR, MIR-S, and MIR-L) of the Infrared Camera (IRC; Onaka e al.\ 2007).

\end{table*}

In a previous work (Mignani et al.\ 2014; hereafter Paper I),  we reported on the positional cross-correlation of a sample of 144 field binary pulsars (i.e. not in Globular Clusters), selected from the Australia Telescope National Facility (ATNF) pulsar  catalogue v1.50 (May 6 2014)\footnote[1]{{\tt www.atnf.csiro.au/research/pulsar/psrcat/index.php?version=1.50} } (Manchester et al.\ 2005),
with source catalogues obtained from  ultraviolet (UV), optical (O), infrared (IR) - UVOIR for short - imaging sky surveys (see Sectn.\ 2.1).
The selection of the starting binary pulsar sample and the cross-correlation with the reference UVOIR source catalogues are discussed in Paper I, to which we refer for details and for an overview of the cross-correlation results.
To summarise, in Paper I we reported on the possible identification of the companion star for three binary pulsars, whereas for other nine binary pulsars, whose companion star had been already identified, we obtained either new UVOIR flux measurements  in at least one of the surveys' spectral bands or multi-epoch photometry.   This allows us to add more information to the study of these nine binary pulsars, which is the subject of our paper.

In particular, we used their serendipitous multi-band follow-up 
to update the spectral energy distribution (SED) of their companion stars,  from the near-UV to the mid-IR.  For some of them, we also used the multi-epoch  catalogue photometry to search for either flux modulations  along the orbital period of the binary system or long-term variability of the companion star  flux. The former are important either to obtain additional identification evidence for the pulsar companion or derive information on the star properties from the comparison of light curves in different bands whereas the latter can unveil variations in the star photosphere/chromosphere activity. The nine binary pulsars discussed in the current work are  both ordinary pulsars and millisecond pulsars (MSPs). Following the classification adopted in Paper I, we defined the former as those with  spin period \Pspin$\ga 0.05$ s and  period derivative $10^{-17}$ s s$^{-1}$ $\lapp  \dot {P}_{\rm s} \lapp 10^{-11}$ s s$^{-1}$ and the latter as those with \Pspin$\lapp10$ ms and $\dot {P}_{\rm s} \lapp 10^{-19}$ s s$^{-1}$ \citep{handbook}.  Six of these nine pulsars are MSPs, of which four have been detected in $\gamma$-rays by {\em Fermi} (Abdollahi et al.\ 2020).

To remark the continuity with 
Paper I, which back then assumed the ATNF pulsar catalogue v1.50 as a reference,  in the current paper  we intentionally decided not to use any later version of this catalogue to expand the above sample of nine binary pulsars. Indeed, this would require to repeat all the cross-correlations with the UVOIR catalogues anew going beyond the immediate goals of this paper, originally aimed to build upon Paper I as its ideal companion paper  (hence, hereafter Paper II), with both conceived as parts of a pilot project aimed at assessing the role of mining serendipitous  (e.g. not always phase connected with the binary orbit) catalogue data in binary pulsar studies.
{\em Although we admit that time-wise Paper II is largely overdue, we deem that, even barring the sample expansion, exploring its results is still important in assessing the usefulness of serendipitous multi-band follow ups and paving the way future similar works by our group.}
Indeed, capitalising on the expertise acquired in the course of this pilot project, a major update of the results presented in both Paper I and Paper II is the goal of  a future work based on the most recent version of the ATNF pulsar catalogue, currently v2.90 (October 6 2026)\footnote[2]{{\tt www.atnf.csiro.au/research/pulsar/psrcat}}, and of the reference UVOIR catalogues, to be updated to include, e.g. the preliminary {\em EUCLID} source catalogue, which we will jointly report in a single paper (Paper III) to set the reference for the next few years.

This manuscript is structured as follows:  in Section\ 2 we briefly summarise the cross-correlations and the flux measurements from Paper I, in Section\ 3 we present the results for the nine pulsars individually. Results are summarised and discussed in Section\ 4.

\section{Multi-band Photometry Database}

\subsection{Source catalogues}

The UVOIR source catalogues used in Paper I were derived from 11 systematic or serendipitous imaging surveys covering different sky fractions and carried out from the near-UV to the mid-IR with both ground and space-based facilities.  

To summarise, these catalogues were: the Galaxy Evolution Explorer ({\em GALEX}; Martin et al.\ 2005) Source Catalogue  (Bianchi et al.\  2011),  the Extreme Ultraviolet Explorer ({\em EUVE}; Bowyer \& Malina 1991) Right Angle Program Source Catalogue (Christian 2002), the  Second \xmm\ optical monitor (OM; Mason et al.\ 2001) Serendipitous Ultraviolet Source Survey catalogue (XMM-SUSS2; Page et al.\ 2012)
the First {\em Swift} UltraViolet and Optical Telescope (UVOT; Roming et al.\ 2005) Serendipitous Source Catalogue (UVOTSSC; Page et al.\ 2014),  the source catalogues from the Sloan Digital Sky Survey (SDSS; York et al.\ 2000),  the UK  Infrared Deep Sky Survey (UKIDSS; Lawrence et al.\ 2007), and the Visible and Infrared Survey Telescope for Astronomy (VISTA; Emerson et al.\ 2006) survey, the {\em Spitzer} Space Telescope (Werner et al.\ 2004) Source List (Teplitz et al.\ 2010),  the {\em Wide-field Infrared Survey Explorer} ({\em WISE}; Wright et al.\ 2010) all-sky survey source catalogue (All{\em Wise}; Cutri et al.\ 2013),  the {\em  AKARI}  Infrared Astronomical Satellite  (Murakami et al.\ 2007)  Point Source Catalogue (Ishihara et al.\ 2010).  
As a reference, in Table \ref{survey} we list the properties of the imaging detectors used by the different telescopes/satellites.
We refer to Section \ 2 of Paper I, and references therein,  for a more comprehensive description of the surveys and their associated source catalogues. 
Whenever a match was found in one of these catalogues
we used data from the Catalina  Surveys (e.g., Drake et al.\ 2009) to look for possible periodic or random variability of the pulsar companion.
Only in some cases we used catalogues from other surveys, such as the GSC2 in the optical (Lasker et al.\  2008)
and 2MASS in the near IR (Skrutskie et al.\ 2006), either as complementary source of information where not available elewhere or for cross-checking the astrometry and photometry. 
The results of the cross-correlation of the ATNF pulsar catalogue with the {\em Gaia} Data Release 2 (DR2; April 25 2018) are reported in Jennings et al.\ (2018).

\subsection{Cross correlation verification}

In Paper I the positional cross-correlation between the binary pulsar list and the source catalogues was carried out assuming a defined {\em association radius}, with a source associated with the pulsar whenever its angular distance was smaller than the association radius.  To account for the different accuracy on the absolute astrometry of different surveys (see Table \ref{survey}) we selected a variable association radius whose value was set to three times the survey nominal astrometry accuracy (see also Section 2.6 of Paper I for details).  We chose such a conservative association radius to account for possible systematics in the survey astrometry, possible offsets between the radio and UVOIR reference frames, and unknown pulsar proper motions. 

In almost all cases, the angular separation between the pulsar and the associated source (Tables 1--8 of Paper I) turned out to be much smaller than the chosen association radius. Nonetheless, we carefully verified  the goodness of the positional match and the identification with the known pulsar companion through chance coincidence probability estimates,  comparison with the known flux measurements of the pulsar companion, inspection of available imaging material, such as published finding charts or archival images, and  relative astrometry. 

As done in Paper I, for each survey catalogue we estimated the chance coincidence probability $P$  from the density $\rho$ of objects in the field, computed within a selected region of $10\arcmin$ radius. This area is large enough to smooth out fluctuations due to possible under and over densities of objects on small angular scales and, thus,  not to under or over estimate the value of $\rho$.  In the case of serendipitous survey catalogues, such as the UVOTSSC and the XMM-SUSS2, we customised the radius of the region to account for both the finite size of the detector and the non-continuous tiling of the sky resulting from different observations. A too large radius would yield to an underestimation of the actual value of $\rho$ and, consequently, to an underestimation of the chance coincidence probability $P$.  Then, we computed $P$ according to the angular separation $\Delta r$ between the pulsar coordinates and those of the matched object in a given catalogue as $P = 1-\exp(-\pi\rho \Delta r^2)$. In all cases, we obtained chance coincidence probabilities as low as $\sim 10^{-4}-10^{-3}$, which are sufficient to rule out spurious matches, especially when complemented by the other cross checks described above. The actual values of $P$ obviously vary on a case-by-case basis, also depending on selection effects such as the different object crowding at different wavelengths, being much lower, e.g. in the near-UV than in the near-IR.  

\begin{table*}
\begin{center}
\caption{Compilation breakdown of the  flux measurements for the nine binary pulsars (see Table 9 of Paper I) with a possible new detection in at least one of the bands of the reference surveys (Table \ref{survey}).  
Some measurements from the UVOTSSC and XMM-SUSS2 catalogues reported in Paper I do not appear in the present compilation (see Sectn.\, 3 for details). Columns report the pulsar name, the identifier of the source matched in a given catalogue, and the magnitude or flux measurements in the survey spectral bands.
All magnitudes are in the AB system, except from the UKIDSS, VISTA, and {\em WISE} ones, which are in the Vega system.  The corresponding AB magnitudes, computed using the equations of Hewett et al.\ (2006),  and Jarrett et al.\ (2011), are reported in brackets. For {\em Spitzer}  sources,  flux values are in $\mu$Jy.
Pulsars for which there are multi-epoch photometry measurements in the Catalina Surveys are at the end of the table, together with their time-averaged Catalina V-band flux. }
\label{all}
\begin{tabular}{lccccccc} \hline \hline
Pulsar & Source ID  & &&Magnitudes/Fluxes&&& \\ \hline \hline
  &   GALEX  &  NUV  & FUV  & & & & \\    
B0655+64 	  & 	 J070037.7+641811 &  	  24.06$\pm$0.26 	&  	& & & &\\
J1012+5307 	
 	                	 & J101233.4+530701 	    & 	20.69$\pm$0.19 & &  & &	& \\
J1023+0038 	& 	 J102347.6+003841 	&    	22.40$\pm$0.13 & & & &	&\\
J1816+4510 	 & J181635.9+451034 & 19.10$\pm$0.09  &	19.82$\pm$0.18	& & & &\\	                \hline 
 &  SDSS & u & g & r & i & z  &  \\ 
J0348+0432	 & J034843.63+043211.3 &21.85$\pm$017 	&  20.71$\pm$0.03& 	 20.61$\pm$0.03 	&  20.69$\pm$0.05 	&  20.38$\pm$0.15  &\\
J1012+5307  & J101233.43+530702.5 	&20.31$\pm$0.05	 & 19.61$\pm$0.01 & 	 19.60$\pm$0.02 	 & 19.71$\pm$0.03 	&  19.84$\pm$0.11 &\\
J1023+0038 	&  J102347.68+003840.9 &17.26$\pm$0.01 	 & 16.68$\pm$0.01 & 	 16.53$\pm$0.01 	&  16.28$\pm$0.01 	&  16.08$\pm$0.01  &\\
J2256$-$1024   & J225656.38$-$102434.4  & 	23.29$\pm$0.71 & 	 22.83$\pm$0.21 & 	 21.60$\pm$0.09 & 	 21.22$\pm$0.11 & 	 20.75$\pm$0.27 & \\ \hline 
 & UKIDSS &  J & H & K  & &   &\\ 
J1023+0038  & 433857550280	&16.63$\pm$0.009 & 16.41$\pm$0.01 	&  & & & \\ 
                     & 	                         &[17.57$\pm$0.009] & [17.79$\pm$0.01] 	&  & & & \\  \hline
 &    VISTA & Z & Y & J & H & K  & \\ 
B1259$-$63 & 515880472129 &9.53$\pm$0.001&12.019$\pm$0.001&8.76$\pm$0.001 &9.58$\pm$0.001 & & \\ 
                    &                          &[10.06$\pm$0.001]&[12.653$\pm$0.001]&[9.69$\pm$0.001] &[10.96$\pm$0.001] & & \\ \hline
               &  XMMOM & UVW1 & UVM2 &UVW2 & U & B & V    \\ 
J0045$-$7319	&  J004535.2$-$731903 & 15.54$\pm$0.01 & & & & &  \\
J1012+5307      &  J101233.4+530702 & & & & 20.23$\pm$0.09 & &  \\
J1023+0038   &  J102347.7+003841&  & & & & 18.10$\pm$0.02 & \\
B1259$-$63     &   J130247.7$-$635008 & & &13.43$\pm$0.01 &  & &  \\  
\hline
        &  SWIFTUVOT& UVW1 & UVM2 &UVW2 & U & B & V  \\ 
B1259$-$63      & J130247.6$-$635008  &   &  13.32$\pm$0.01     &   &   & &      \\ 
J1810+1744 & J181037.2+174437           &                               &                                 &                            &  22.69$\pm$0.15 & & \\
J1816+4510  & J181635.9+451034          &                                 & 18.97$\pm$0.04     & 19.03$\pm$0.05 & 18.50$\pm$0.02 &  &  \\ \hline
              &  WISEA & W1 & W2 & W3 & W4 & &\\ 
J1023+0038      & J102347.69+003840.8 &15.853$\pm$0.048 & 16.079$\pm$0.187 & &  & &\\
                          &                                      &[18.266$\pm$0.048] & [19.398$\pm$0.187] & &  & &\\
B1259$-$63        & J130247.65$-$635008.6 &6.783$\pm$0.038 & 6.462$\pm$0.021 & 5.637$\pm$0.015& 4.741$\pm$0.030 & & \\ 
                           &                                         &[9.466$\pm$0.038] & [9.781$\pm$0.021] & [10.879$\pm$0.015]& [11.345$\pm$0.030] & & \\ \hline
             &  SSTSL2 & 3.6$\mu$m & 4.5$\mu$m & 5.8$\mu$m & 8.0$\mu$m & & \\ 
J0045$-$7319     &   J004535.23$-$731903.2  &70.41$\pm$1.29  & 46.04$\pm$1.62 & & & & \\         \hline          
& CSSS & V & & & & &  \\
J0348+0432 & J034843.7+043209 &  20.32 & & & & & \\
J1012+5307 & J101233.5+530702 &  19.74 & & & & &\\
J1023+0038 & J102347.7+003841&  17.28 & & & & &\\
J1816+4510 & J181635.9+451036 &  18.50 & & & &  & \\
J2256$-$1024 & J225656.2$-$102436 &  21.45 & & & & &\\ \hline \hline
    \end{tabular}
\end{center}
\end{table*}

\subsection{Cross correlation Compilation}

In Table \ref{all}, we produce a compilation of the flux information for the nine binary pulsars (see Table 9 of Paper I) for which the cross-correlations with the different UVOIR source catalogues (Table \ref{survey}) 
provided a new detection of the companion star in at least one  of the surveys' spectral bands and/or multi-epoch flux information from the Catalina Surveys. Magnitudes and flux measurements are compiled from Tables 1--8 of Paper I.  Time-averaged V-band flux measurements from the Catalina Surveys are also included, for completeness. However, being extrapolated to the V band and not calibrated directly\footnote[3]{{\tt http://nesssi.cacr.caltech.edu/DataRelease/FAQ2.html\#reference}}, we will not use the Catalina fluxes for the SED characterisation of the pulsar companion stars. However, we will use them for quantitative comparisons with other fluxes, e.g. as an evidence of identification of a companion star in the Catalina surveys.
As it can be seen,  for no pulsar we obtained new multi-band photometry from the correlations with the {\em EUVE} and {\em AKARI} source catalogues. 

Soon after Paper I was published, the UVOTSSC catalogue was released to the community\footnote[4]{{\tt http://www.ucl.ac.uk/mssl/astro/space\_missions/swift/uvotssc}}. Therefore, we checked the results of the cross-correlations reported in Paper I, which used a beta-release of the UVOTSSC (Yershov 2014),  against the first official catalogue release (version 1.1). Table \ref{all} reports the revised matches and the updated magnitude values.  We could not recover the match with one of the pulsars (PSR\, J0045$-$7319) reported in Table\, 3 of Paper I, whereas for another one (PSR\, B1259$-$63) we could not recover the magnitude values in the U, UW1 and UVW2 pass bands. 
For the latter pulsar, magnitudes in the U,B,V and UVW1, UVM2 bands from the XMM-SUSS2 catalogue were filtered out  after a sanity check (see, Section\, 3 for details).

We note that different surveys in  Table \ref{survey} use different photometric systems. For instance, magnitudes from the UKIDSS, VISTA, and {\em WISE} surveys are given in the Vega system, whereas those of the other surveys are in the AB system (Oke 1974).   
As a reference, we converted magnitudes from the Vega to  the AB system using, per each band, the corresponding transformations available in the literature. For both UKIDSS and VISTA, we used the equations of  Hewett et al.\ (2006),  whereas for {\em WISE} we used the equations of Jarrett et al.\ (2011).  The  computed AB magnitudes  for the UKIDSS, VISTA, and {\em WISE} sources are listed in Table \ref{all} together with the native catalogue Vega magnitudes.

The coordinates,  spin period ($P_{\rm s}$), period derivative ($\dot{P_{\rm s}}$), and inferred values of spin-down age ($\tau$), dipolar magnetic field ($B$), rotational energy loss ($\dot{E}$) of the nine pulsars listed in Table \ref{all} are summarised in Table \ref{psr}. Their orbital parameters, orbital period P$_{\rm b}$, eccentricity $e$, and inclination angle $i$,  and the expected companion mass and type are summarised in Table \ref{orb}.   Among these nine pulsars, six are MSPs (J1012+5307, J1023+0038, B1259$-$63, J1810+1744, J1816+4510, J2256$-$1024) of which the last five have been detected in $\gamma$-rays by {\em Fermi} (Abdo et al.\ 2013). 

\begin{table*}
\begin{center}
\caption{Coordinates, spin-down parameters  (period $P_{\rm s}$ and period derivative $\dot{P_{\rm s}}$) of the pulsars listed in Table \ref{all}, together with the inferred values of the spin-down age ($\tau$), dipolar magnetic field ($B$) and rotational energy loss ($\dot{E}$), as derived from the ATNF pulsar catalogue.  For pulsars with measured radio proper motions (PSR\,  J0348+0432, J1012+5307, J1023+0038, B1259$-$63, and J1816+4510) the values of $\dot{P_{\rm s}}$, as well as the derived values of $\tau$, $B$ and $\dot{E}$, have been corrected for the Shlovskhii effect (Shlovskhii 1970). 
 }
\label{psr}
\begin{tabular}{lllccccc} \hline
Pulsar 				&$\alpha_{J2000} $	& $\delta_{J2000}$&    P$_{\rm s}$	&    $\dot{P_{\rm s}}$ 	& $\tau$ & B & $\dot{E}$ \\ 
                                          &    $^{(hms)}$   &      $^{(\circ ~'~")}$   &   (ms) &  (10$^{-18}$s s$^{-1}$)    &  ($10^{9}$ yr) & ($10^{9}$ G)  & ($10^{34}$ erg cm$^{-2} s^{-1}$)  \\ \hline
J0045$-$7319   	&  00 45 35.16(7)      	& $-$73 19 03.0(2)     	 	  &  926.276 	& 4463.2 	 	 &   0.003 	 &2060 	&    0.022 \\
J0348+0432	         &  03 48 43.639                 & +04 32 11.4580       &                39.1         &     0.235                &    2.64                     &    3.07         &     0.0155      \\
B0655+64     	& 07 00 37.802(2)    	 	 &+64 18 11.20(2)    	 	  &195.671 	&    0.663 	  	  &  4.68 	 &  11.5 	&    0.00035 \\
J1012+5307   	 &10 12 33.4339(1)  	  	 &+53 07 02.574(1)  	 	   & 5.256 	    &0.0114 	     	    &7.30 	   & 0.248	  &  0.31 \\
J1023+0038   	 &10 23 47.687196(2) 	  	 &+00 38 40.84551(4)  & 1.688 	    &0.0102 	     	    &2.63 	    &0.133 	    &8.37 \\
B1259$-$63      & 13 02 47.65(1)                        &$-$63 50 08.7(1)           & 4.77        & 2280        &0.00032  & 334       & 82.61 \\
J1810+1744       & 18 10 37.28(1)                        & +17 44 37.38(7)      & 1.66      &      -                 &           -      &    -          &  -             \\
J1816+4510   	 &18 16 35.9314(2)   	  	 &+45 10 33.864(2)   	 	   & 3.2	    &0.041 	&   1.240 	    &0.367 	&    4.9 \\
J2256$-$1024   	 &22 56 56.39(1)     	  	 &$-$10 24 34.37(12)    		    &2.29 	    & - & 	-    & -	 &  -  \\ \hline
\end{tabular}
\end{center}
\end{table*}

\begin{table*}
\begin{center}
\caption{Orbital parameters (orbital period P$_{\rm b}$ and eccentricity $e$) of the pulsars listed in  Table \ref{psr} as derived from the ATNF pulsar data base (second and third columns), together with the estimated or assumed orbital inclinational angle ($i$), the assumed or measured pulsar mass ($M_{\rm P}$) and the measured or lower limit on the companion mass ($M_{\rm C}$) inferred from the system mass function. Only for PSR\, J0045$-$7319,  J0348+0432, and J1012+5307 it was possible to determine the system mass ratio from the measurement of Doppler shifts in the companion spectrum.  Proposed classifications of the companion star are given in the following column, either main sequence (MS) or white dwarf (WD). The classification of the system according to the expected companion star irradiation (BW=black widow; RB= red back) is 
also indicated. Cases not confirmed by spectroscopy are highlighted in italic. References for the companion mass measurement and classification are given in the last column.}
\label{orb}
\begin{tabular}{lcccccll} \hline
Pulsar 				& P$_{\rm b}$  	& $e$                 &$i$              & $M_{\rm P}$     & $M_{\rm C}$     &Class.     & Refs. \\ 
                                          &     (d)                     &                         &($^{\circ})$  & ($M_{\odot}$)  & ($M_{\odot}$)    &     & \\ \hline
J0045$-$7319   	        &   51.17                 & 0.807949         & 44(3)  & 1.4                   & 8.8(18)      &  MS  & Kaspi et al.\ (1994); Bell et al.\  (1995) \\ 
J0348+0432		         &    0.102                         &   2.4$\times10^{-6}$              & 40.2(6)    &    2.01(4)               &         0.172(3)                                          & WD &Antoniadis et al.\  (2013).\\
B0655+64     		        &   1.03 	                & 2.7(25)$\times10^{-6}$ &         90         &    1.4                      &       $>$0.81&  WD                      &  Camilo (1997) \\
J1012+5307   	                 & 0.60 	                & 1.0(6)$\times10^{-6}$   &      52(4)           &  1.64(20)            &0.156(20) &  WD    &  Callanan et al.\ (1998); van Kerkwijk et al.\ (1996)  \\ 
J1023+0038   	                & 0.19 	                & $\la 2\times10^{-5}$ &90   & 1.4        & $>$0.138                   & MS,RB         & Breton et al.\ (2013); Takata et al.\ (2013) \\
B1259$-$63                      & 1236.72                 & 0.8698869              &         &              &                                    &    MS   & Johnston et al.\ (1994) \\
J1810+1744                      & 0.15                       &    -                           & 90          &  1.4                   &      $>$0.045 &   {\it MS,BW}    &   Breton et al.\ (2013)\\
J1816+4510   	                & 0.36                     &       -                         & 90        &          1.4          & $>$0.162     &     WD & Kaplan et al.\  (2013) \\
J2256$-$1024   	      & 0.21                      &        -                        & 90$^{*}$ & 1.4           & $>$0.030       &     {\it MS,BW}   & Breton et al.\ (2013) \\  \hline 
\end{tabular}
\end{center}
$^{*}$ $i=68^{\circ}\pm11^{\circ}$ from the optical light curve (Breton et al.\ 2013)
\end{table*}

\section{Results}

In the following sections we describe the results of the cross-correlations for these nine pulsars, compare catalogue flux measurements with those  available in the literature, and describe those obtained in new spectral bands.  Whenever possible, we verified the flux measurements from the re-analysis of the original survey images. Moreover,  for five pulsars we also present the light curves obtained from the Catalina Surveys. In all cases, we folded the light curves using the known orbital period and the orbital ephemerides  of the binary system given in the ATNF data base. By definition, we assumed the epoch of periastron to set the zero of the orbital phase. For the light curve analysis, we used the GRaphical Analyzer of TIme Series (GRATIS) code. This is a software developed at the Bologna Observatory by P. Montegriffo, which uses both the Lomb-Scargle periodogram (Lomb 1976; Scargle 1982) and the best fit of the data with a truncated Fourier series (e.g., Barning 1963) and has been extensively applied to search for periodicity in variable stars (e.g., Sollima et al.\ 2010; Fiorentino et al.\ 2010; Muraveva et al.\ 2014).

\subsection{J0045$-$7319} 
\label{J0045}

The optical companion of  the pulsar J0045$-$7319 (P$_{\rm s}$=0.926 s) in the Small Magellanic Clouds (SMC) was identified by Kaspi et al.\ (1994) and spectroscopically classified as a massive ($M_{\rm C} \sim 10 M_{\odot}$) and bright B1 main sequence (MS) star (B=16.03; V=16.19), consistent with the mass estimates obtained from the system mass function (Table \ref{orb}). The identification was later confirmed by Bell et al.\ (1995), based on radial velocity variations in the star spectrum. No other observations at UVOIR wavelengths have been reported ever since. The binary pulsar might have been detected in X-rays once during a survey of the SMC with the {\em Rossi X-ray Timing Explorer}  (Galache et al.\ 2008) but no X-ray pulsations were detected. 

Owing to its brightness, the pulsar companion is also detected in the 2MASS 
(J=16.599$\pm$0.156, H=16.797,  K$_s$=16.292) 
and GSC2.3 (R$_F$=16.05; B$_J$=15.61) catalogues (with all magnitudes in the Vega system), not included in this work.  
The pulsar companion had also a match in the beta-release of the UVOTSSC (Paper I; Table\, 3). However, as anticipated in Paper I, the source was detected close to the edge of the UVOT detector and has not been included in the official  catalogue release.
In the XMM-SUSS2 catalogue, we identified the pulsar companion with source XMMOM\, J004535.2$-$731903 ($\Delta r = 0\farcs50$), of magnitude $m_{\rm UVW1}=15.54\pm0.01$. This would be the first identification of the star in the near-UV band to date.   Unfortunately, 
the star is detected in only one OM observation
and the field is not covered by the Catalina Surveys. Thus, we could not search for modulations at the orbital period (P$_{\rm b}$=51.17 d). These were searched for but not found in data collected by the Optical Gravitational Lensing Experiment (OGLE)  that, instead, showed evidence for a super-orbital period at 2654$\pm$ 298 d (Rajoelimanana et al.\  2011). The star is also identified in the {\em Spitzer} Source List at an angular separation $\Delta r = 0\farcs41$ from the radio coordinates, with fluxes of  70.41$\pm$1.36 and  46.04$\pm$1.65   $\mu$Jy at 3.6$\mu$m and 4.5$\mu$m, respectively. Like in the near-UV, this would be the first measurement of the star flux in the mid-IR. 

\begin{figure}
\centering
{\includegraphics[width=8cm,bb=18 24 592 324,clip=]{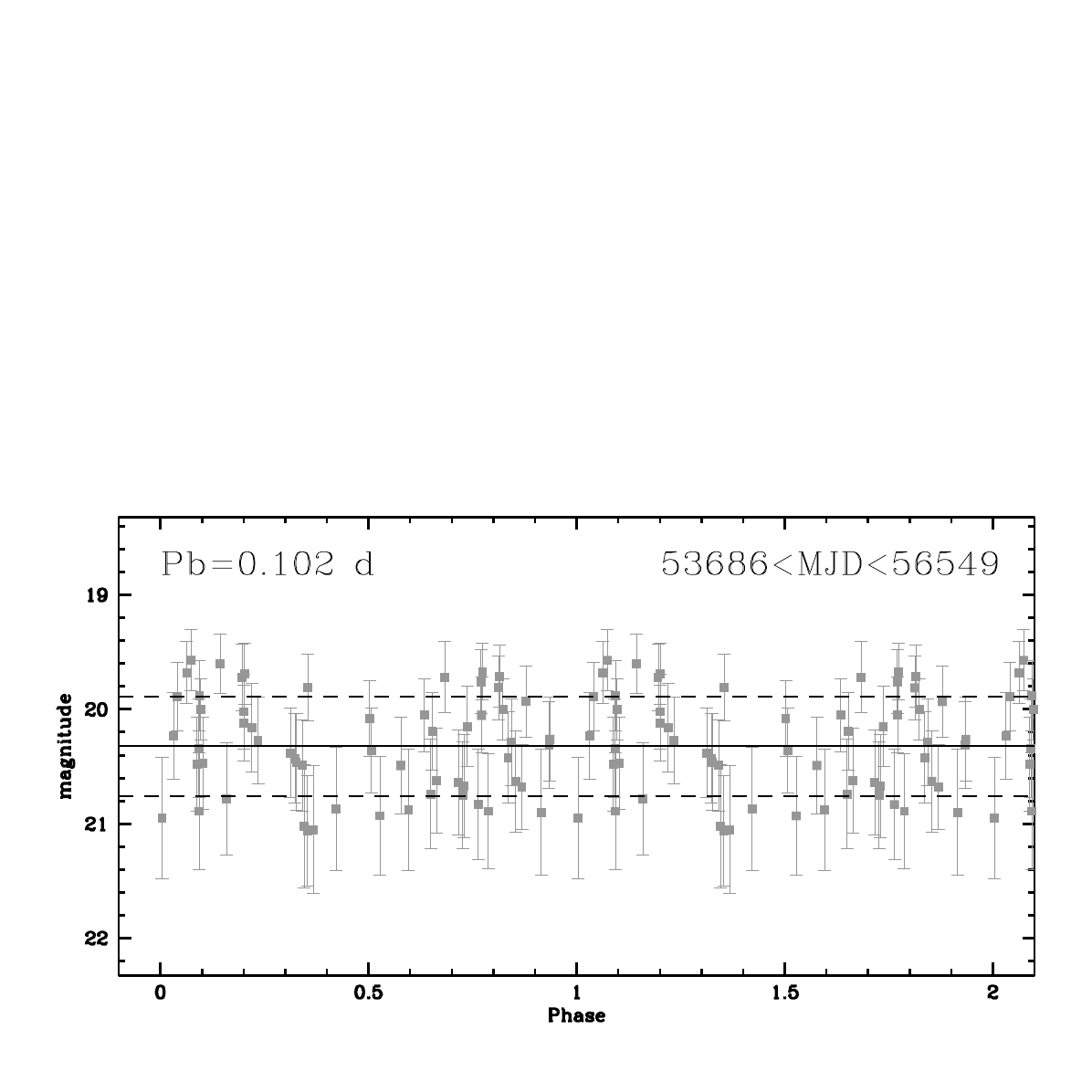}}
\caption{\label{J0348_lc} Catalina V-band light curves of PSR\, J0348+0432 folded around the orbital period P$_{\rm b}$ of the binary system. Two orbital cycles are shown for clarity. The solid line corresponds to the mean of the flux measurements whereas the dashed lines correspond to the $1 \sigma$ scatter around the mean. The epoch range spanned by the Catalina flux measurements is reported on the top right of the panel.
}
\end{figure}

\subsection{J0348+0432} 
\label{J0348}

With a spin period P$_{\rm s}$=39.1 ms  and a low magnetic field $B=3.07 \times 10^{9}$ G, PSR\, J0348+0432 is a mildly-recycled pulsar (Lynch et al.\ 2013).  We found a positional match ($\Delta r = 0\farcs19$) between PSR\, J0348+0432 and SDSS\, J034843.63+043211.3  ($u$=21.85$\pm$017; $g$=20.71$\pm$0.03; $r$=20.61$\pm$0.03; $i$= 20.69$\pm$0.05; $z$=20.38$\pm$0.15). This is the same source that Lynch et al.\ (2013) identified as the pulsar companion in the same SDSS data, and that was subsequently classified as a He-core low-mass WD based upon optical spectroscopy (Antoniadis et al.\  2013). 
No evidence for orbital  modulation (P$_{\rm b}$=0.102 d) from the WD companion to PSR\, J0348+0432 was found by Antoniadis et al.\  (2013).

We possibly identified the star  ($\Delta r = 1\farcs49$)  with the Catalina source CSS\, J034843.7+043209 of magnitude V=20.32, consistent with the SDSS photometry. The Catalina dataset consists of flux measurements obtained in 58 epochs, distributed over a time span of almost 8 years (MJD=53686--56549).    We downloaded the flux measurements from the Catalina archive and folded them around the orbital period of the binary system.  The Catalina light curve only shows a very marginal hint of a sinusoidal modulation
(Fig.\ref{J0348_lc}),
which  however, is not statistically significant. 
Therefore, this pulsar is not considered in the following discussion.

\subsection{B0655+64} 
\label{B0655}

The companion star of PSR\, B0655+64  (P$_{\rm s}$=0.195 s) was detected by  Kulkarni (1986) as a carbon-oxygen (CO) WD (R=22.1; V=22.2), spectroscopically identified as such by van Kerkwijk \& Kulkarni (1995),   with $T_{\rm eff} =$ 6000--9000 K. 

 We identified the pulsar companion in the \gal\  Guest Investigator Survey (GII) catalogue (GALEX\, J070037.7+641811; $\Delta r = 0\farcs49$) with NUV magnitude of 24.06$\pm$0.26. The goodness of the positional match, within the $1 \sigma$ uncertainty on the absolute \gal\ astrometry  (see Table \ref{survey})
suggests that we likely detected the PSR\, B0655+64 companion star in the near-UV for the first time. 
Unfortunately, the pulsar companion is too faint to be detected in the Catalina surveys. Deeper observations are needed to search for a  possible orbital flux modulation.

\subsection{J1012+5307} 
\label{J1012}

Soon after its discovery in radio, the optical companion to the MSP J1012+5307  (P$_{\rm s}$=5.256 ms) was identified in the Digitised Sky Survey images (Nicastro et al.\ 1995). Multi-colour photometry (Lorimer et al.\ 1995) gave V=19.58$\pm$0.02; B=19.78$\pm$0.04; R=19.49$\pm$0.03; I=19.32$\pm$0.03 for the companion star, which was spectroscopically classified as an He WD (van Kerkwijk et al.\ 1996)  with $T_{\rm eff} =  8550 \pm 25 $ K and log$~g =6.75 \pm 0.07$ (see also,  Callanan et al.\ 1998). PSR\, J1012+5307 has been detected as a $\gamma$-ray pulsar by {\em Fermi} (Abdo et al.\ 2013) and  as an X-ray pulsar by \xmm\ (Webb et al.\ 2004). 

\begin{figure}
\centering
{\includegraphics[width=8cm,bb=18 24 592 324,clip=]{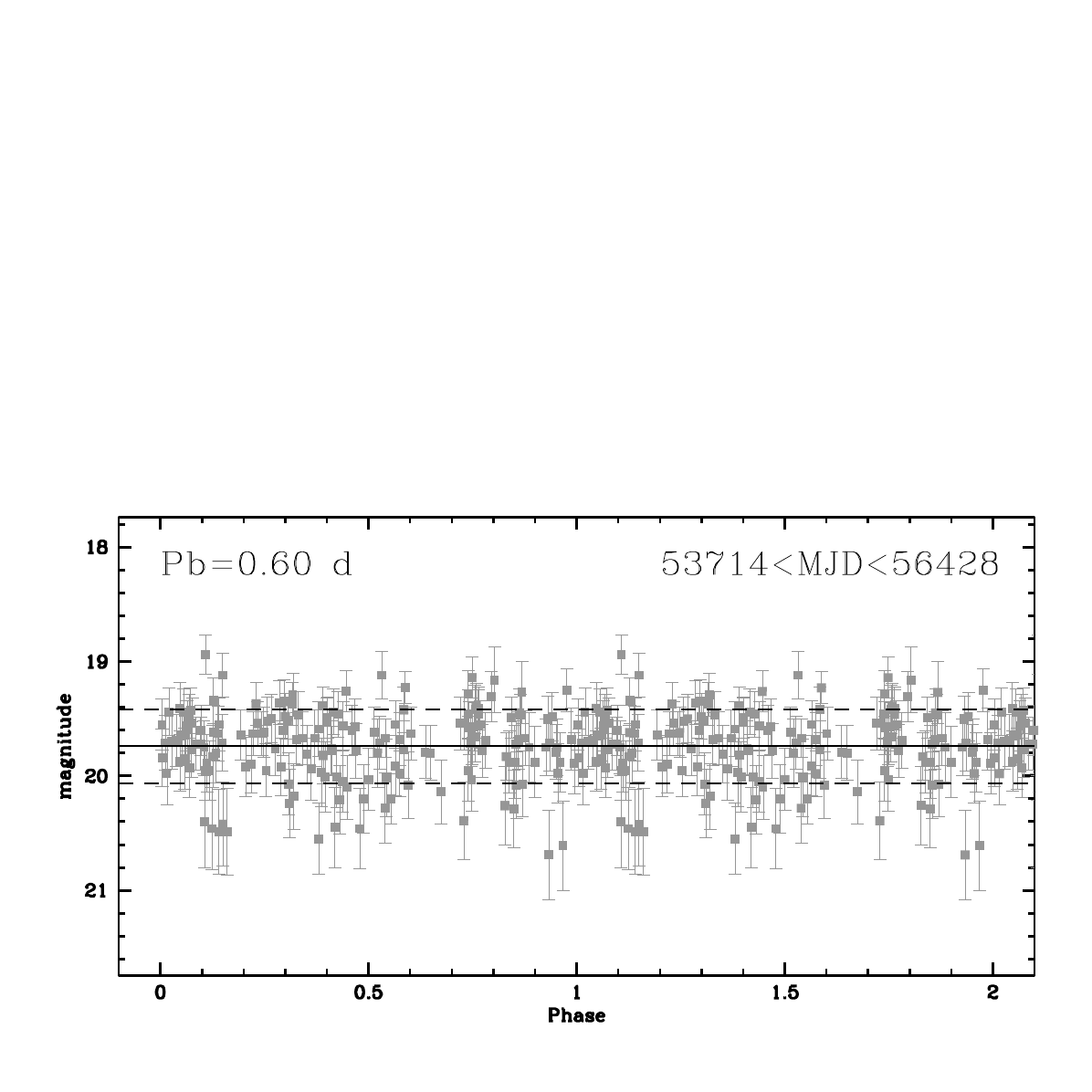}}
\caption{\label{J1012_lc} Catalina V-band light curves of PSR\, J1012+5307 folded around the orbital period ($P_{\rm b}$) of the binary system. Two orbital cycles are shown for clarity
}
\end{figure}

We identified the pulsar companion with source SDSS\, J101233.43+530702.5 ($\Delta r = 0\farcs09$) with magnitudes u=20.31$\pm$0.05, g=19.61$\pm$0.01, r=19.60$\pm$0.02, i=19.71$\pm$0.03, z=19.84$\pm$0.11 and in the XMM-SUSS2 catalogue with source XMMOM\, J101233.4+530702  ($\Delta r = 0\farcs30$), detected only in one OM observation (U=20.23$\pm$0.09). The SDSS photometry is consistent with that of Lorimer et al.\ (1995), once colour transformations between the Sloan and the Johnson bands and magnitude systems (AB and Vega) are taken into account.  In addition, we obtained unpublished multi-band photometry information in both the u and the z band, with the SDSS u-band magnitude consistent with the XMM-SUSS2 U-band one.  Moreover, we identified the pulsar companion in both the \gal\  GII  ($\Delta r = 0\farcs46$) and Medium Imaging Survey (MIS)  ($\Delta r = 0\farcs58$) catalogues (GALEX\, J101233.3+5307020), with NUV magnitudes  of 21.19$\pm$0.07 and 21.29$\pm$0.07, respectively. Accounting for the accuracy of the \gal\ absolute astrometry ($\delta r = 0\farcs5$; Table \ref{survey}), this source probably coincides with  GALEX\, J101233.4+530701 ($\Delta r = 1\farcs01$) detected in the All-sky Imaging Survey  (AIS) with NUV magnitude 20.69$\pm$0.19, consistent (within $3\sigma$) with the measurements in the GII and MIS surveys. Thus, we probably detected the PSR\, J1012+5307 companion star in the near-UV for the first time.  
The light curve of the PSR\, J1012+5307 companion star has not been measured yet.
We found a likely positional association for the star  in the Catalina Surveys  ($\Delta r = 0\farcs23$) with source CSS\, J101233.5+530702 (V=19.74). Unfortunately,  despite of the large number of data points  in the Catalina light curve (Fig.\ref{J1012_lc}), distributed over 165 epochs and spanning about 7.5 years (MJD=53714--56428), we found no evidence of possible flux modulations at the orbital period (P$_{\rm b}$=0.60 d) of the binary system, with most of the points within $\pm 1 \sigma$ from the mean.  Owing to the relatively large error bars  and the data point scatter, we can only rule out modulations of amplitude larger than $\sim$0.6 magnitudes ($1 \sigma$).

\subsection{J1023+0038} 
\label{J1023}

\begin{figure}
\centering
{\includegraphics[width=8cm,bb=18 24 592 324,clip=]{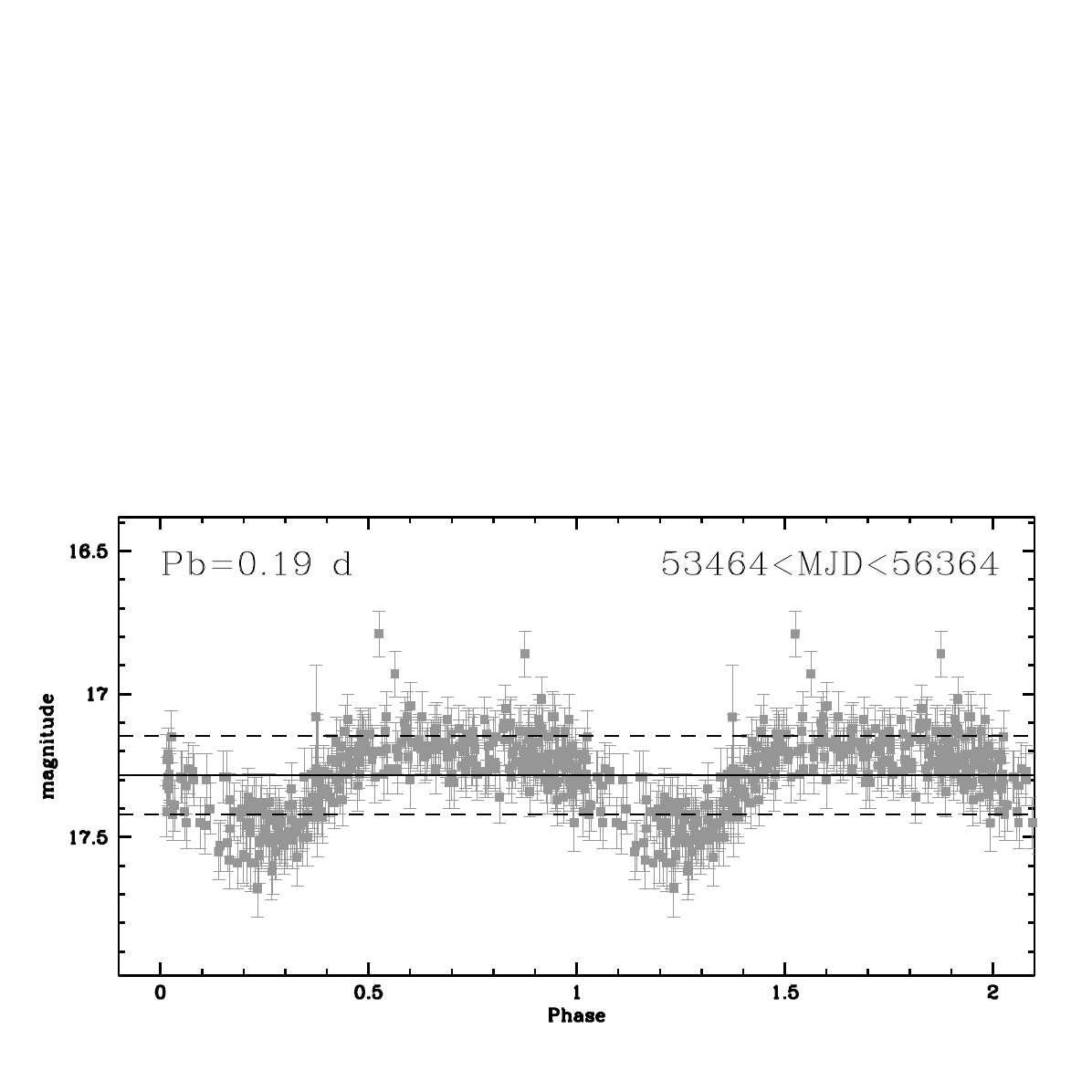}}
{\includegraphics[width=8cm,bb=18 254 592 554,clip=]{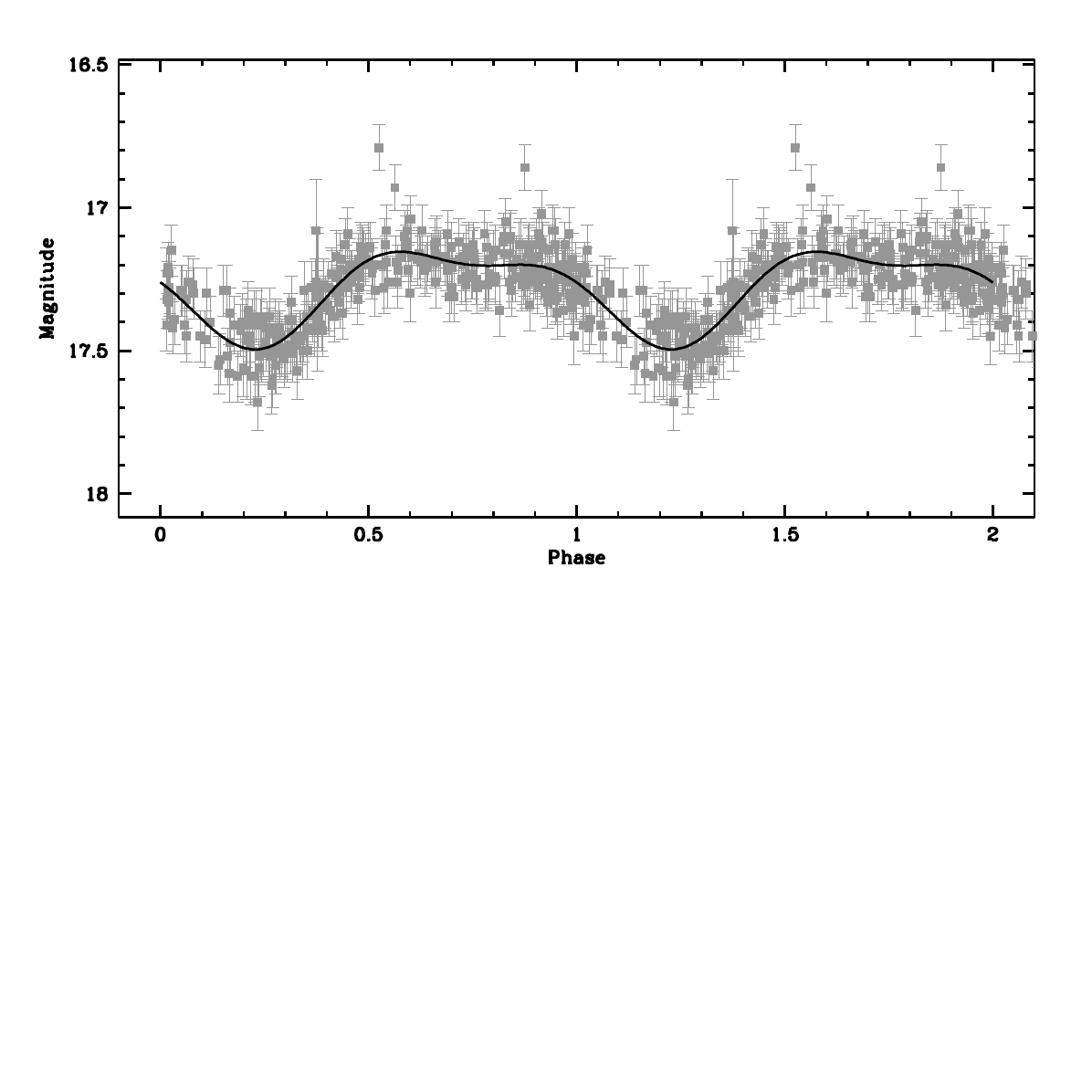}}
\caption{\label{J1023_lc} (top) Catalina V-band light curves of PSR\, J1023+0038 folded around the orbital period ($P_{\rm b}$) of the binary system. (bottom) Same but with the best-fit model computed by GRATIS superimposed (solid line). Two orbital cycles are shown for clarity.
}
\end{figure}

The companion star to the MSP J1023+0038  (P$_{\rm s}$=1.688 ms; Archibald et al.\ 2009) was identified in the optical prior to the pulsar radio discovery, during follow-up observations of the unidentified radio source FIRST\, J102347.6+003841 (Bond et al.\ 2002) and also detected in the first data release (DR1) of the SDSS (Szkody et al.\ 2003). Modulations of the optical flux at a 0.198 d period, only later found to be the orbital period of the binary pulsar (Archibald et al.\ 2009), were discovered by Woudt et al.\ (2004) and confirmed by Thorstensen \& Armstrong (2005).  Optical pulsations from J1023+0038 have been detected in a sub-luminous accretion-powered state by Papitto et al.\ (2013). We identified the PSR\, J1023+0038 companion   in the Catalina Surveys ($\Delta r = 0\farcs46$) with source CSS\, J102347.7+003841 (V=17.28) and clearly detected the orbital modulation also in the Catalina data, which we show here for the first time (Fig. \ref{J1023_lc}, top). The Catalina light curve is broadly consistent with those of Woudt et al.\ (2004) and Thorstensen \& Armstrong (2005), also obtained in the V band, with a similar flattish maximum and amplitude of $\approx$ 0.5 magnitudes.  We note that the peak of the Catalina light curve is at orbital phase $\Phi_{\rm b}$$\sim$0.7 instead of 0.4 (Woudt et al.\ 2004; Thorstensen \& Armstrong 2005) but this is purely due to a different definition of the reference epoch.  The modulation is better seen by fitting the folded light curve with GRATIS (Fig. \ref{J1023_lc}, bottom), which hints at a  possible smooth double hump superimposed on the light curve maximum. However, given the photometric uncertainty on the data points, we conclude that this double hump is probably not statistically significant.

We identified the pulsar companion  both in the SDSS (J102347.68+003840.9; u=17.26$\pm$0.01; g=16.68$\pm$0.01; r=16.53$\pm$0.01; i=16.28$\pm$0.01; z=16.08$\pm$0.01) and  XMM-SUSS2 (XMMOM\, J102347.7+003841; B=18.03$\pm$0.02) catalogues ($\Delta r = 0\farcs16$).  We note that the SDSS fluxes in DR10 are different from those in DR1 (Szkody et al.\ 2003), probably due to the average of photometry measurements obtained over different epochs. PSR\, J1023+0038 is detected as an X and $\gamma$-ray source and, together with PSR\, J1824$-$2452I (Papitto et al.\ 2013), PSR\,  J1227$-$4853 (Roy et al.\ 2014) and, possibly, 1RXS\, J154439.4$-$112820 (Bogdanov \&  Halpern 2015), is one objects  in between MSPs and LMXBs, which show evidence for discontinuous matter accretion phases from their companions and consequent transitions from radio active/X-ray quiescent to radio quiescent/X-ray active states (hence called "transitional" MSPs). For this reason, PSR\, J1023+0038  is known to be variable at all wavelengths, both in flux and spectrum (e.g., Patruno et al.\ 2014; Stappers et al.\ 2014; Takata et al.\ 2014; Tendulkar et al.\ 2014), according to the accretion state.  Thus, it is, in principle, not surprising that optical photometry taken at different epochs reflects such variability.

We also detected the companion star in the near-IR in the UKIDSS Large Area Survey  ($\Delta r = 0\farcs07$), with magnitudes J=16.63$\pm$0.009 and H=16.41$\pm$0.01 (Vega).  These values are slightly fainter with respect to both the magnitudes listed in 2MASS (J=16.30$\pm$0.10; H=15.69$\pm$0.13, K$_{s}$=15.9$\pm$0.3) and those reported in Wang et al.\ (2013) (J=16.260$\pm$0.024; H=15.964$\pm$0.025).  
This difference is larger than expected from colour equations between different filters and, once again, it might be ascribed to the system variability.  However, the 2MASS observations and those of Wang et al.\ (2013) were taken on 2000 February 6 and 2012 May 10, respectively, whereas the UKIDSS ones were taken on 2008 March 23 and 2009 December 13. This means that all measurements were taken in an epoch range when the IR emission of the system would be dominated by the companion star and not by an accretion disk (Wang et al.\ 2013), hence would be relatively stable. One possibility to explain the photometry difference would be that the UKIDSS observations were taken at different orbital phases with respect to those of 2MASS and Wang et al.\ (2013), assuming that there is an orbital modulation in the IR as in the optical.  At variance with Wang et al.\ (2013) who claimed   detections at 3.6 and 4.5 $\mu$m, we found no match for PSR\, J1023+0038 in the {\em Spitzer} Source List, probably due to the catalogue assembling criteria. 
 In the mid-IR, however, we identified PSR\, J1023+0038 in the All{\em WISE} Source Catalogue (WISEA\, J102347.69+003840.8; $\Delta r = 0\farcs04$), with magnitudes $m_{\rm W1}$=15.853$\pm$0.048 and $m_{\rm W2}$=16.079$\pm$0.187 (Vega), consistent with those reported in Wang et al.\ (2013) which were obtained from a previous catalogue release. 
  
 Finally, we also identified PSR\, J1023+0038 both in the \gal\ MIS ($\Delta r = 0\farcs54$) and AIS ($\Delta r = 1\farcs11$)  surveys (GALEX\, J102347.6+003841), with NUV magnitudes of 22.40$\pm$0.13 and 22.04$\pm$0.66, respectively.  Time wise, the {\em GALEX} ones, obtained back in March 30, 2006, are the first detections  of the pulsar binary system in the near-UV. 
 Thereafter, Patruno et al.\ (2014) observed  PSR\, J1023+0038 with the \sw/UVOT  on June 12 and October 18, 2013 and measured UW1 Vega magnitudes of $\approx$19.8 (see also Rivera Sandoval et al.\ 2018) and 16.32$\pm$0.05, respectively, with the latter indicating a re-brightening of the system following  the turn-off of the radio emission (Stappers et al.\ 2014) and the onset of the X-ray active phase (Kong 2013), with the formation of an accretion disk. 
Similar UVW1 magnitude values (16.90--16.12) were measured in follow-up UVOT observations (Takata et al.\ 2014).  A detailed description of an extended UVOT monitoring campaign of PSR\, J1023+0038  is given in Coti Zelati et al.\ (2014). A new near-UV detection of PSR\, J1023+0038 was obtained by  Papitto et al.\  (2019)  with the {\it Swift}/UVOT and the   UVW1 filter on May 23, 2017 during an accretion disk state, with magnitude $16.05 \pm 0.03$ in the Vega system. These measurements were not recovered in our cross-correlations since the version of the UVOTSSC used in Paper I only includes data from the first five years since the launch of {\em Swift} (2005--2010). 
The  \gal\  March 30, 2006 NUV flux measurements are contemporaneous to a non-accretion disk state between 2001 and 2007 (Archibald et al.\ 2009), with the pulsar being radio-active and X-ray quiescent in that epoch range. For this reason,  the \gal\ NUV flux can be used to build the stellar component to the binary system SED since it is virtually unaffected by the contribution of an X-ray irradiated accretion disk, whereas the contribution of the pulsar itself to the total NUV emission is uncertain\footnote[5]{Near UV time-resolved spectroscopy observations (1570--3180 \AA) with the {\it Hubble Space Telescope} in  April 2014,  June 2017 and 2021 detected pulsed emission from PSR\, J1023+0038 (Miraval Zanon et al.\ 2022). However, the source occurred to be in a  sub-luminous accretion disk state and the emission was attributed to non-thermal 
radiation in a region just outside the light cylinder close to the disk shock with the pulsar wind.}.
      
\subsection{B1259$-$63} 
\label{B1259}

PSR\, B1259$-$63 is a well-known MSP (P$_{\rm s}$=4.77 ms) in an eccentric and long orbit  (e=0.86; P$_{\rm b}$=1236.72 d) around the very bright  (V=10.15) MS companion (LS\, 2883; Stephenson \& Sanduleak 1971) and detected from the X-rays up to the very high-energy $\gamma$-rays (e.g., Abramowski et al.\ 2013). The star is classified as a B2e (Johnston et al.\ 1994) 
and, more recently, as an O9.5 V by Negueruela et al.\ (2011). 

The PSR\, B1259$-$63 companion  is detected  in both the XMM-SUSS2 (XMMOM\, J130247.7$-$635008; $\Delta r = 0\farcs50$) and UVOTSSC (SWIFTUVOT\, J130247.6$-$635008; $\Delta r = 0\farcs48$) catalogues.
However, in the optical bands
the star  is in the central enhanced area of high OM background,  it suffers of severe {\em mod8} noise, and  is run through a significant readout streak. Moreover, it is fairly saturated in both the optical 
and UV bands. 
Therefore, on a second analysis, we discarded the XMM-SUSS2 photometry apart from the UVM2 one ($m_{\rm UVW2}$=13.43$\pm$0.01)
The UVOTSSC magnitude is $m_{\rm UVM2}$=13.32$\pm$0.01. Thus,  the PSR\, B1259$-$63 companion star would be detected in the near-UV for the first time.
Detections of the pulsar companion also in the UVW1, UVW2 and U bands 
were reported in the beta-release of the UVOTSSC  (Paper I; Table\, 3). However,
we found that the star is at the edge of a region affected by bright smoke rings in the UVW1 and U-band images, whereas it is slightly saturated on the UVW2 one. For these reasons, the star magnitudes in these bands were not included in the official UVOTSSC release.  
Accounting for the pass band difference, the OM UVW2 and UVOT UVM2 band magnitudes  are compatible with each other. 
Thus,  the PSR\, B1259$-$63 companion star would be detected in the near-UV for the first time.
The single OM UVW2  and UVOT UVM2 observations of PSR\, B1259$-$63 were taken when the companion was approaching the periastron and we could not spot any obvious evidence of variability, 
consistently with what reported by other  observations. Unfortunately, the pulsar field is not covered by the Catalina data.
The star is also detected in the VISTA VVV survey in the Z, Y, J, H filters
but inspection of the VIRCAM images shows that it is saturated. 
In the mid-IR, the PSR\, B1259$-$63 companion is detected in the All{\em WISE} catalogue ($\Delta  r =0\farcs15$) and, possibly, in the {\em  AKARI} All-Sky Survey Point Source Catalogue ($\Delta r = 2\farcs6$), although the association is uncertain because the angular separation is about twice as large as the association radius.  For this reason, this match is not listed in Table \ref{all}. The All{\em WISE} magnitudes are $m_{\rm W1}$=6.783$\pm$0.038; $m_{\rm W2}$= 6.462$\pm$0.021; $m_{\rm W3}$=5.637$\pm$0.015;  $m_{\rm W4}$=4.741$\pm$0.030 (Vega), whereas the {\em  AKARI}  flux densities would be 0.2933$\pm$0.0142 and 0.1339$\pm$0.0256  Jy in the  S9W (9.0 $\mu$m) and the L18W (18.0 $\mu$m) bands, respectively i.e. not directly comparable to the All{\em WISE} ones.

\subsection{J1810+1744} 
\label{J1810}

The MSP (P$_{\rm s}$=1.7 ms) J1810+1744  was also discovered in a radio follow-up of unidentified {\em Fermi} sources (Hessels et al.\ 2011) and subsequently detected as a $\gamma$-ray pulsar (Abdo et al.\ 2013). The companion star of PSR\, J1810+1744 was identified in the optical by Breton et al.\  (2013), who also measured flux modulations at the orbital period.  These have also been seen in multi-band light curves (BVR) of the companion star recently obtained by Schroeder \& Halpern (2014). No spectral classification of the PSR\, J1810+1744 companion is available yet. 

We identified the companion star in the UVOTSSC catalogue with source SWIFTUVOT\, J181037.2+174437  ($\Delta r = 0\farcs27$) and measured its magnitude in the U band for the first time (U=22.69$\pm$0.15).
).  
We searched the Catalina data base but, unfortunately, the pulsar field is not covered by the survey.

\begin{figure}
\centering
{\includegraphics[width=8cm,bb=18 24 592 324,clip=]{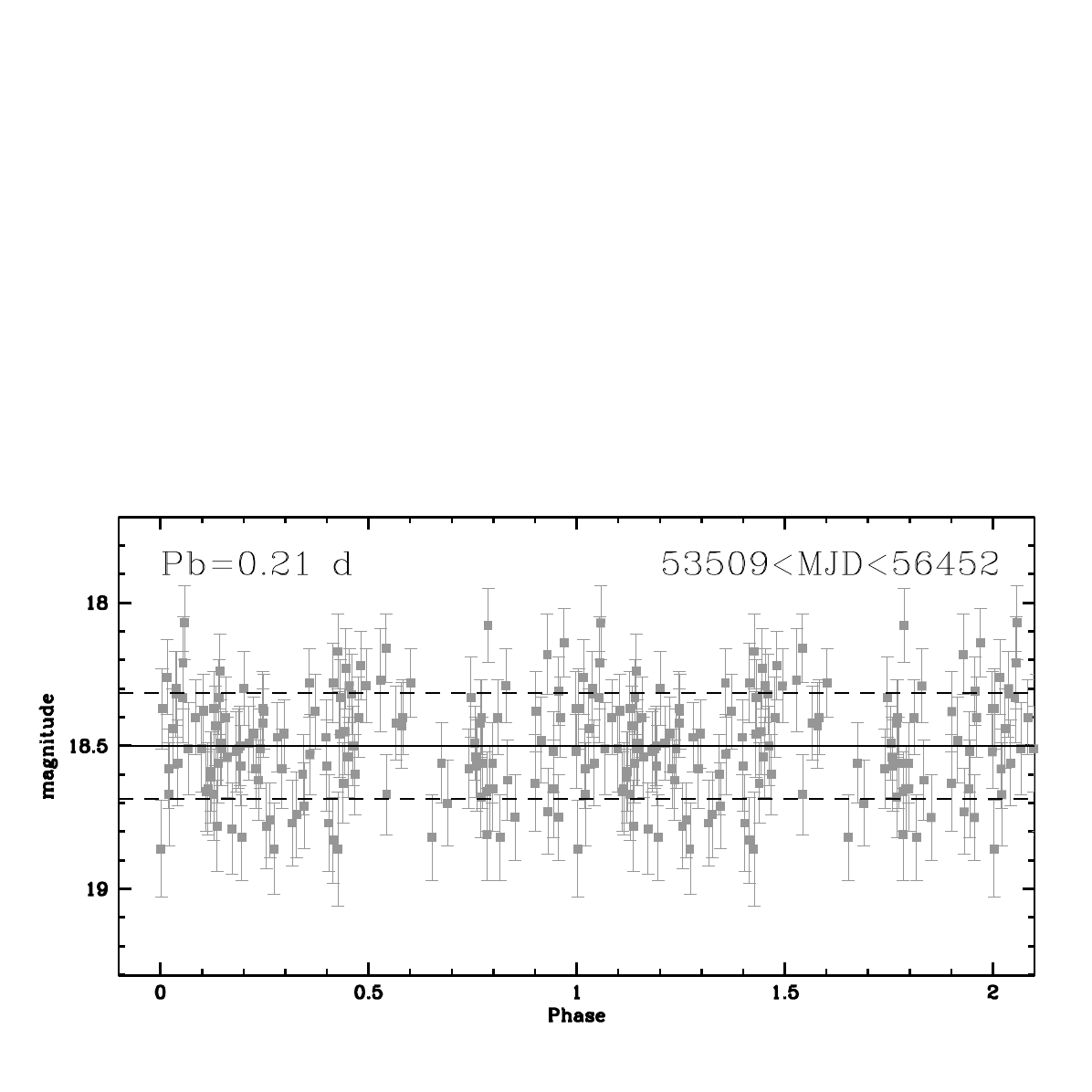}}
{\includegraphics[width=8cm,bb=18 24 592 324,clip=]{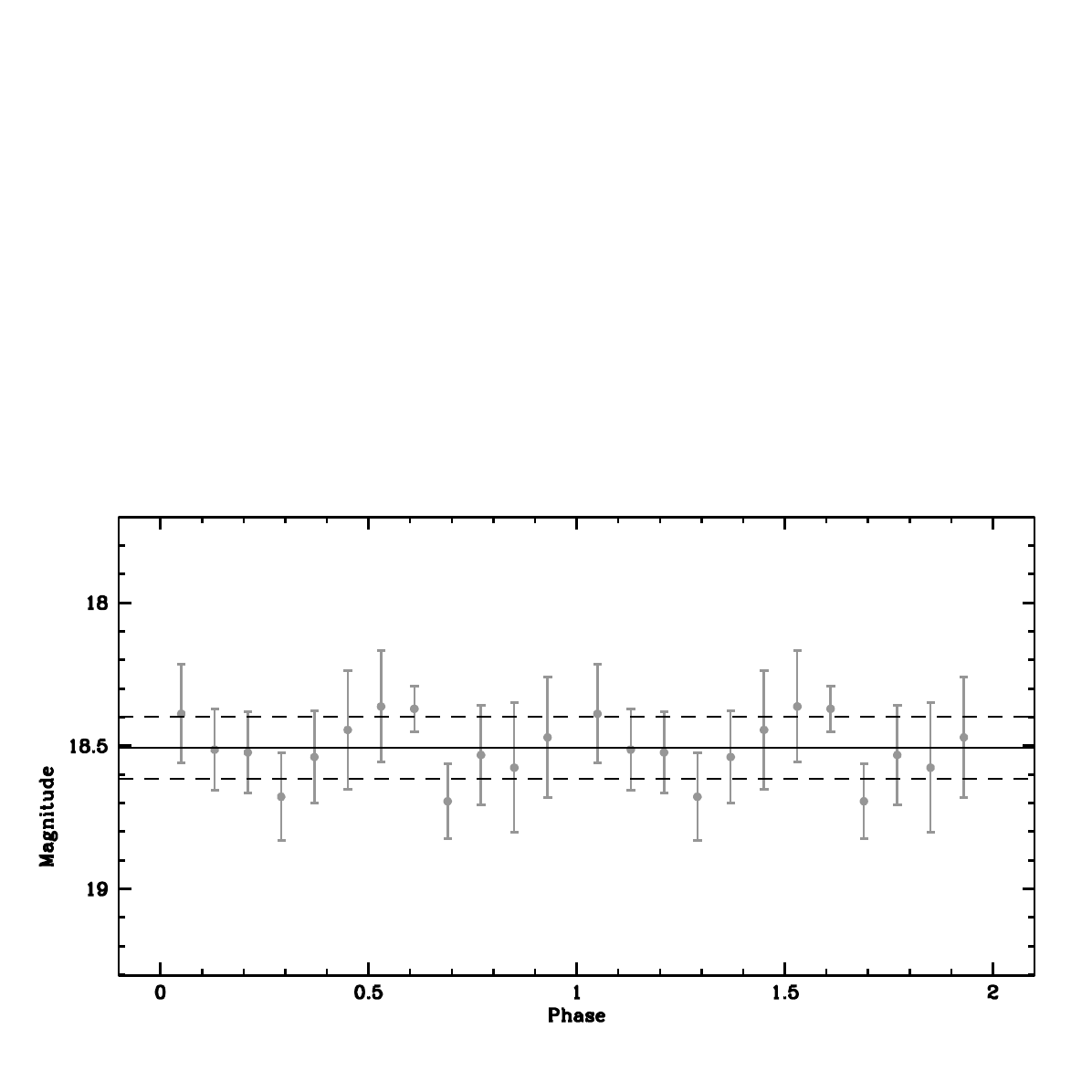}}
{\includegraphics[width=8cm,bb=18 264 592 564, clip=]{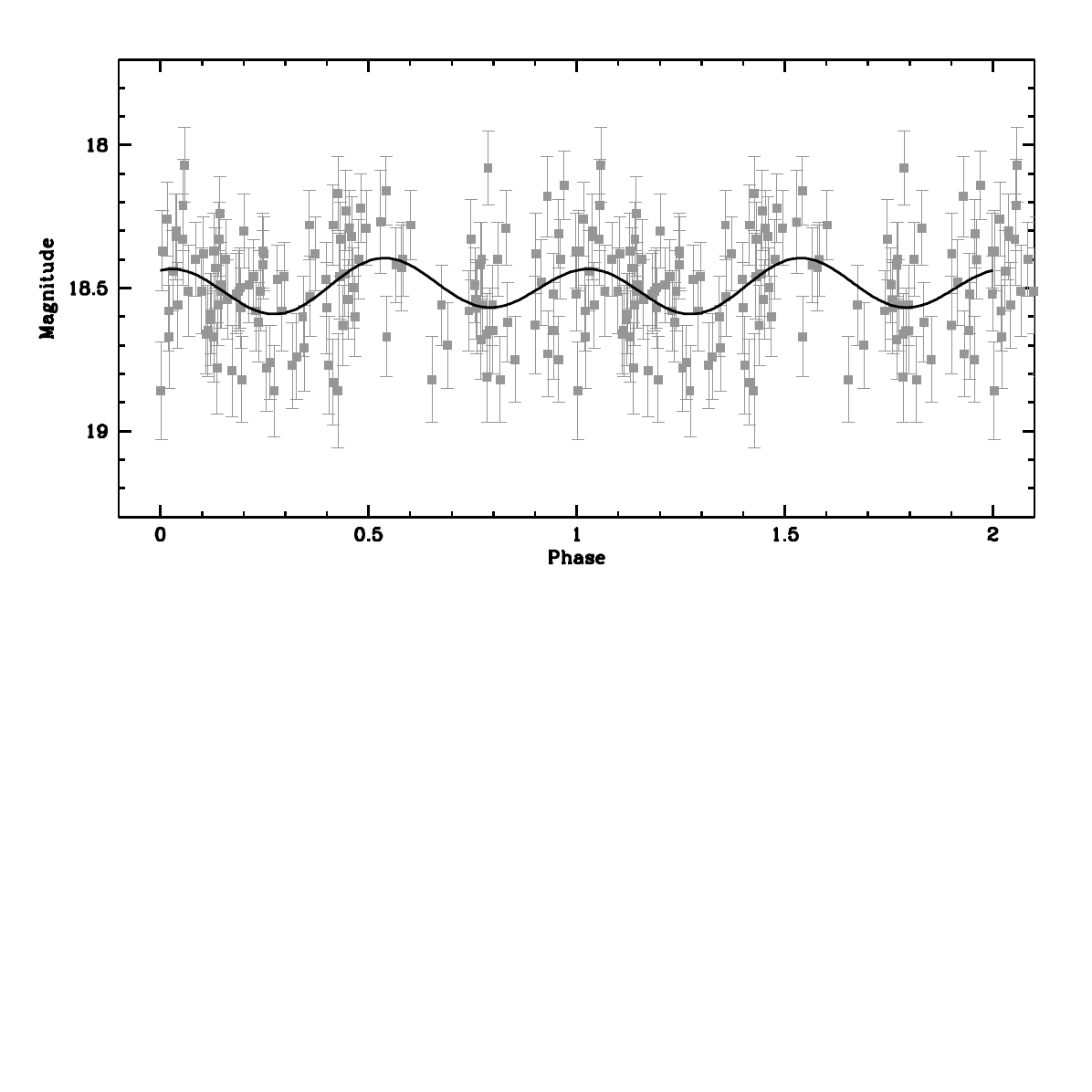}}
{\includegraphics[width=8cm,bb=18 24 592 324,clip=]{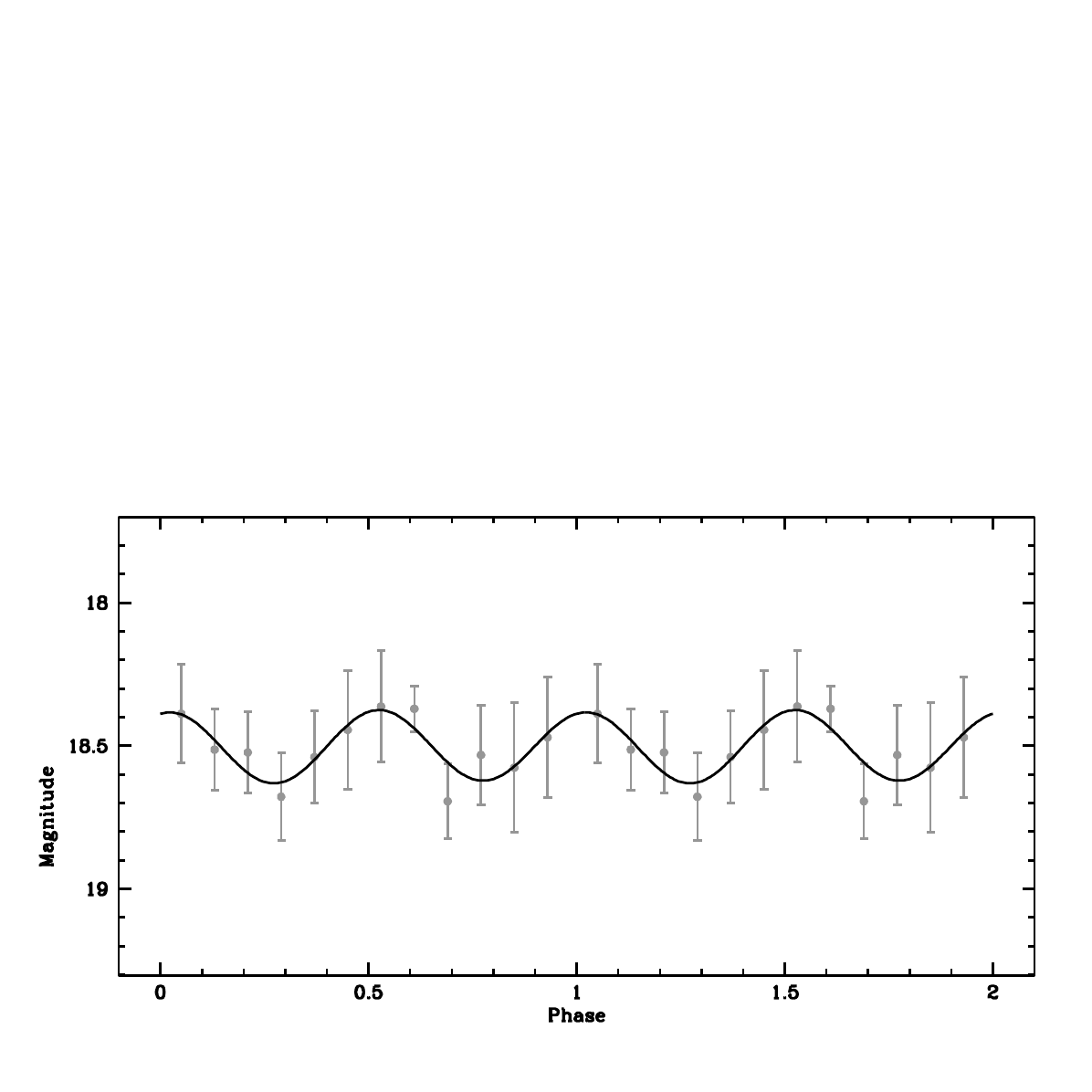}}
\caption{\label{J1816_lc}  (top two panels) Catalina V-band light curves of PSR\, J1816+4510 folded around the orbital period ($P_{\rm b}$) of the binary system and binned in phase by 0.08.  
(bottom two panels) Same, but with the best-fit model computed by GRATIS superimposed (solid line).   Two orbital cycles are shown for clarity.
}
\end{figure}

\subsection{J1816+4510} 
\label{J1816}

PSR\, J1816+4510 is a MSP (P$_{\rm s}$=3.2 ms) discovered in the error box of the unidentified {\em Fermi} source 2FGL\,  J1816.5+4511 and also identified as a $\gamma$-ray pulsar (Kaplan et al.\ 2012).  Based upon ground and space-based observations, including those from the \sw/UVOT and \gal, Kaplan et al.\ (2012) discovered the companion star of the pulsar in the optical/near-UV.
Follow-up spectroscopy observations have been obtained by Kaplan et al.\ (2013), which confirmed the identification through radial velocity measurements and suggested a proto-white dwarf/sdB star classification for the companion, with $T_{\rm eff} = 16000 \pm 500 $ K and log$~g =4.9 \pm 0.3$.  

The measured \gal\ NUV and FUV magnitudes that we derived from the catalogue source GALEX\, J181635.9+451034  ($\Delta r =0\farcs42$),  19.10$\pm$0.09  and 19.82$\pm$0.18, are consistent with those of Kaplan et al.\ (2012). Their UVOT U-band magnitude is also consistent with that obtained from the UVOTSSC catalogue  for source SWIFTUVOT\, J181635.9+451034 ($\Delta r =0\farcs12$; U=18.50$\pm$0.02). In addition, however, we identified the pulsar companion  in the UVM2 and UVW2 bands (OBS ID 00041440002 and 00041440001), with magnitudes $m_{\rm UVM2}$=18.97$\pm$0.04 and $m_{\rm UVW2}$=19.03$\pm$0.05. 
The four UVM2-band exposures of the PSR\, J1816+4510 field  
do not indicate any obvious variability for SWIFTUVOT\, J181635.9+451034, owing to the small fraction of the orbital period  (P$_{\rm b}$=0.361 d) covered by the exposure sequence (less than 5 hours).  Flux modulations at the orbital period were also searched for in the UVOT U-band light curve by Kaplan et al.\ (2012) but were not found. A possible peak ($\sim$ 0.3 magnitudes) near orbital phase 0.75 was seen in the Catalina light curve, tough, after rebinning the data points.   Modulations at the orbital were also searched in more recent observations (Schroeder \& Halpern 2014) but, owing to the low number of flux measurements (six), they were not found. 

To investigate the existence of peak in the light curve reported by Kaplan et al.\ (2012), we downloaded the Catalina data from the archive. Two Catalina sources are matched with PSR\, J1815+4510, CSS\, J181635.9+45103 and CSS\, J181635.9+451036, which have the same average brightness of V$\sim$18.5 and whose coordinates differ by $\approx$ 2\arcsec.  As noted by Kaplan et al.\ (2012), these two sources are likely the same object identified at slightly different positions by the Catalina pipeline, possibly because of random problems with the image astrometry. Inspection of the Digitised Sky Survey (DSS) images, which have a spatial resolution similar to the Catalina ones, confirms this conclusion. The Catalina photometry covers 130 epochs distributed over an $\sim$ 8 year time span (MJD=53509--56452). Eight of these data points, with magnitudes V$\sim$16.6--16.9, likely correspond to a brighter star southeast of the pulsar position, once again mis-identified with the pulsar companion because of a problem in the Catalina astrometry, and we removed them (see also Kaplan et al.\ 2012). We then folded the remaining data around the orbital period of the binary pulsar using the published radio ephemerides. 
Although the light curve is quite noisy (Fig. \ref{J1816_lc}, top panel), the data points show some fluctuations around the mean (reduced $\chi^{2}=1.7$), which might hint at an orbital modulation.  

For a better visualisation, we re-sampled the light curve in different phase bins, from 0.08 to 0.12, chosen to have both a statistically adequate and a more or less constant number of data points per bin (about 10).  As an example, we show in Fig. \ref{J1816_lc} (second panel from top) the light curve obtained with a phase bin of 0.08 (12 data points).  The rebinned  light curve  hints  more clearly at two possible maxima ($\Phi_{\rm b}\sim $0 and $\Phi_{\rm b}\sim$0.5) and a minimum ($\Phi_{\rm b}\sim$ 0.3), whereas the uneven phase coverage of the unbinned data points around orbital phase $\Phi_{\rm b}$=0.7 makes it is difficult to recognise the possible presence of a second minimum.   
To better model this possible flux modulation (reduced $\chi^{2}=0.76$), we fitted both the unbinned and binned Catalina light curves using GRATIS.  In both cases, the light curve  fluctuations are compatible with a sinusoidal modulation (reduced $\chi^{2}=1.57$ and 0.31, respectively) of amplitude $\approx$ 0.2 magnitudes, with two maxima and two minima per cycle. These are separated in phase by $\sim 0.25$ (Fig. \ref{J1816_lc}, bottom two panels), as expected in case of a modulation due to distortion of the companion star. Thus, our light curve is different from that of Kaplan et al.\ (2012),  possibly owing to a different binning  of the data points and to a different orbital phase definition.
However, the Catalina data do not allow us to statistically discriminate between our fitted model and a flat distribution.  Thus, although we see a hint of an orbital modulation in the data, we cannot firmly confirm it without new, more precise, observations.  The  fact that  the PSR\, J1815+4510 companion is not in the Catalina catalogue of periodic variable sources of Drake et al.\ (2014) does not necessarily argue against the presence of such modulation in the data since the non unambiguous association of the star with a Catalina source (see above) might have hampered their variability analysis.

\begin{figure}
\centering
{\includegraphics[width=8cm,bb=18 24 592 324,clip=]{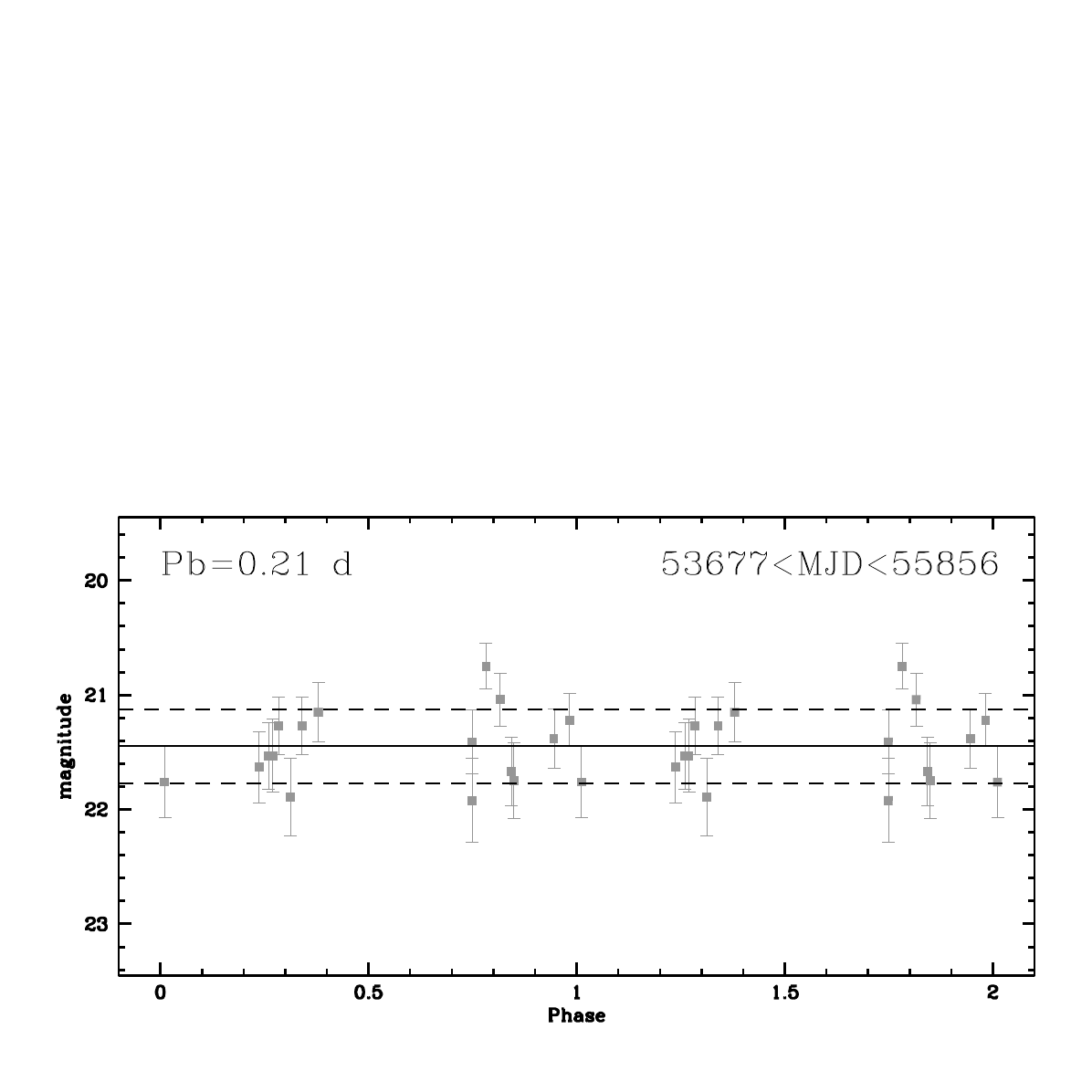}}
\caption{\label{J2256_lc} Catalina V-band light curves of PSR\, J2256$-$1024.
}
\end{figure}

\begin{table*}
\begin{center}
\caption{Summary of the available UVOIR spectral flux information for all pulsars listed in Table \ref{all}.  Detections in the surveys' spectral bands are marked by the band names. Detections presented in this work  for the first time are highlighted in bold. The pulsar identification in the multi-epoch data  from the Catalina Surveys and the presence of a clear, or possible, flux modulation at the orbital period of the binary system is flagged as P, whereas a detection but with no statistically evidence of flux modulation is flagged as D. }
\label{summ}
\begin{tabular}{lcccccccccc} \hline
Pulsar 				&GALEX	& UKIDSS	 &   SDSS	&    UVOT	& VISTA  	& OM & Spitzer & WISE  & Catalina \\ \hline
J0045$-$7319	        &		&		&		&	&	 	& {\bf UVW1}	        & {\bf 3.6,4.5} & &   \\
J0348+0432		        &		&		&	ugriz	&		&		& 	  	        & & & {\bf D} \\
B0655+64	        &     {\bf NUV}	&		&		&		&	 	&      		&  & &  \\
J1012+5307	        &     {\bf NUV}	&		&	{\bf u}gri{\bf z}	&		&	 	&  U   	& & &  {\bf D} \\
J1023+0038	        &     {\bf NUV}	&	JH	&	ugriz	&		&	 	&  B   	& &W1,W2 & {\bf P}\\
B1259$-$63		        &	      	&		&		&	{\bf UVM2}	&	 	
& {\bf UVW2} & &{\bf W1--W4} &  \\  
J1810+1744              &                       &               &              &  {\bf U}          &              &               &  & &  \\
J1816+4510            &     NUV,FUV	&		&		&	{\bf UVM2},{\bf UVM2},U	&	 	& 			& & &  {\bf P}\\  
J2256$-$1024		        &		&		&	{\bf u}g{\bf r}i{\bf z}	&		&	 	&      		& & & {\bf D} \\  \hline 
\end{tabular}
\end{center}
\end{table*}

\subsection{J2256$-$1024}
\label{J2256}

The MSP J2256$-$1024 (P$_{\rm s}$=2.29 ms) is also one of the  $\gamma$-ray pulsars detected by {\em Fermi} (Abdo et al.\ 2013).  PSR\, J2256$-$1024 is  also one of the pulsars for which Breton et al.\ (2013) identified the optical companion and measured the light curve. No spectral classification for the companion star is available yet.  We found a very good positional match ($\Delta r = 0\farcs03$) between PSR\, J2256$-$1024 and source SDSS\, J225656.38$-$102434.4, of magnitudes  u=23.29$\pm$0.71, g=22.83$\pm$0.21, r=21.60$\pm$0.09, i=21.22$\pm$0.11, z=20.75$\pm$0.27. 

Our SDSS $g$ and $i$-band magnitudes of the companion star are consistent with those 
of Breton et al.\ (2013),  whereas we obtained
measurements in the  $u$, $r$, and $z$ bands for the first time.  The companion star might have also been  detected  in the Catalina Surveys  (MLS\, J225656.2$-$102436; V=21.45), although the offset from the pulsar position ($\Delta r = 3\farcs11$) is much larger than the nominal $3\sigma$ uncertainty on the Catalina absolute astrometry (1\farcs5; see Table \ref{survey}).  Nonetheless, the 
case of  PSR\, J1816+4510 
(Section 3.8) shows that the Catalina coordinates can be affected by random offsets  which can be of about this order of magnitude.  However, assuming that MLS\, J225656.2$-$102436 is the PSR\, J2256$-$1024 companion, with only 16 epochs the available photometry measurements 
are too sparse and unevenly distributed along the orbital phase (with a large gap around $\Phi_{\rm b} = 0.5$) to look for variability at the 0.21 d orbital period (Fig. \ref{J2256_lc}).

\section{Summary}

We obtained new multi-band photometry information for nine binary pulsars for which the companion star had been already identified (Table 9 of Paper I).
Furthermore, for five of these pulsars  we also obtained multi-epoch flux information  from the Catalina Surveys, although only for PSR\, J1023+0038 it allowed us to clearly detect the orbital flux modulation (Section 3.5). A summary of the available UVOIR flux information for these nine pulsars (Table \ref{all}) is given in Table \ref{summ} where, for each of them, we highlight a new detection in a given survey spectral band. 

For  four  pulsars,  we obtained new detections in either  near-UV (PSR\, J1816+4510) or optical (PSR\, J0348+0432, J1012+5307, J1810+1744, J2256$-$1024) spectral bands.  
For four pulsars (PSR\, J0045$-$7319, B0655+64, J1012+5307, J1023+0038) the cross-match with sources in the \gal, UVOTSSC, or XMM-SUSS2 catalogues yielded the first-time near UV detection of the pulsar companion. Particularly important is the contribution of \gal, which yielded three of the new near-UV detections. For two pulsars (PSR\, J0045$-$7319, B1259$-$63) the cross-match with either one or both of the {\em Spitzer} and   {\em WISE}
catalogues yielded the first detection of the companion star in the mid-IR.   
Thus, the largest number of detections in new spectral bands are obtained in the near-UV and in the mid-IR
Finally, for five pulsars (PSR\, J0045$-$7319, J1012+5307, B1259$-$63, J1816+4510, and J2256$-$1024) we obtained new photometry measurements in at least two bands.

In the following we present our results for two different cases. For pulsars whose companion star classification has already been proposed through optical spectroscopy observations, i.e. PSR\, J0045$-$7319, B0655+64, J1012+5307, J1023+0038, B1259$-$63, and J1816+4510  
(see Table \ref{orb} and references therein), our updated photometry measurements serve as a base both to verify consistency with the published SEDs and improve or expand the SED characterisation at longer and shorter wavelengths. For pulsars whose companion star has only tentatively classified through broad-band photometry, i.e. PSR\, J1810+1744 and J2256$-$1024 (Breton et al.\ 2013), 
our new photometry measurements serve as a base to further verify the proposed classification\footnote[6]{Both analyses will be reported in the Appendix, to be added to the submitted version of the manuscript.}.

We summarise the results for the two groups of pulsars, i.e. those with or without spectral classifications, in two following subsections separately.

\subsection{Pulsars with spectroscopy-based companion star classification}

There are six pulsars in this group, PSR\, J0045$-$7319, B0655+64, J1012+5307, J1023+0038, B1259$-$63, and J1816+4510, 
which we discuss below on a case by case basis. Spectral parameters have been assumed from the literature and distance and reddening are those reported in Table \ref{dist}.
PSR\, J0045$-$7319 -- We obtained the first flux measurements of the companion star in both the near-UV with the OM   ($m_{\rm UVW1}= 15.54\pm0.01$), mid-IR with {\em Spitzer} in the 3.6$\mu$m and 4.5$\mu$m bands (70.41$\pm$1.29 $\mu$Jy, 46.04$\pm$1.62 $\mu$Jy), and incidentally in the  near-IR with 2MASS (J=16.599$\pm$0.156, H=16.797, K$_{\rm  s}$=16.292), not included in Table \ref{summ}. This allows us to expand the available spectral coverage, so far limited to the optical range (Kaspi et al.\ 1994), to be compared with a template spectrum for a $(8.8 \pm 1.8) M_{\odot}$  B1-type star (Kaspi et al.\ 1994).

PSR\, B0655+64 -- We obtained the first near-UV detection with {\em GALEX} ($m_{\rm NUV}$=24.06$\pm$0.26), to be  compared with the simulated spectrum of a CO WD ($T_{\rm eff} =$6000--9000 K), as proposed from the optical spectrum of the companion star obtained by   van Kerkwijk \& Kulkarni (1995).

PSR\, J1012+5307 -- We obtained the first u and z-band magnitude measurements from the SDSS (20.31$\pm$0.05, 19.84$\pm$0.11, respectively) and the first near-UV measurement from {\em GALEX} ($m_{\rm NUV}$=20.69$\pm$0.1), to be compared with the companion spectrum which is described by a temperature $T_{\rm eff} = 8550\pm 25$ K and log g = 6.75$\pm$0.07 (van Kerkwijk et al.\ 1996).

PSR\, J1023+0038 --  {\em GALEX} provided the first--in--time NUV flux measurements ($m_{\rm NUV}$=22.40$\pm$0.13 and 22.04$\pm$0.66), to be compared with the optical/UV SED of the binary system at different accretion states (e.g., Tendulkar et al.\ 2014). The Catalina survey yielded independently the companion star light curve in the V band (see Sectn. \ref{J1023}), which is consistent with those in the literature. 

PSR\, B1259$-$63 --  UVOT and OM observations provided magnitude measurements in both the UVM2 (13.32$\pm$0.01) and UVW2 (13.43$\pm$0.01) bands, respectively. the first in the near UV.  The first fluxes in the MIR  were obtained from {\em WISE} in the W1 to W4 bands (magnitudes 6.783$\pm$0.038, 6.462$\pm$0.021, 5.637$\pm$0.015, 4.741$\pm$0.030), respectively. The optical spectrum of the companion star is that of an O9.5 V class star (Negueruela et al. 2011).

PSR\, J1816+4510 -- The pulsar companion is detected in the UVOT UVM2 and UVW2 bands (magnitudes 18.97$\pm$0.04, 19.03$\pm$0.05, respectively), to be compared with 
its spectrum (Kaplan et al.\ 2013) which is described by  a temperature $T_{\rm eff} = 16000 \pm 500$ K and log g = 4.9 $\pm$0.3.

\subsection{Pulsars with unconfirmed companion star classification}

There are only two pulsars in this group, which  we discuss separately. 

PSR\, J1810+1744 --
The only detection that we obtained through our cross-correlations is in the UVOT U band ($22.69\pm0.15$). Very little can be said on the nature of the companion star without colour information other than from the UVOT data alone, where  only a uvw1-band detection is available (Breton et al.\ 2013). We searched the star in both the 2MASS and the GSC2.3 catalogues and only found a match with the latter, with optical magnitudes of R$_{\rm F}=20.14 \pm 0.63$ and B$_{\rm J}=21.02 \pm 0.43$, roughly compatible with the time-averaged R and B-band ones of Schroeder \& Halpern (2014) once filter differences are taken into account.

PSR\, J2256$-$1024 -- We detected the pulsar companion in the SDSS u, r, z bands (magnitudes 23.29$\pm$0.71, 21.60$\pm$0.09, 20.75$\pm$0.27) for the first time, whereas detection in the g and i bands (22.83$\pm$0.21, 21.22$\pm$0.11)  have been reported elsewhere (Breton et al.\ 2013). The CMD analysis of the SDSS data can provide clues on the nature of the companion star.

To be taken as a reference for forthcoming analyses, we have compiled from the literature the most recent values of distance and interstellar reddening.
The distance, inferred either from the dispersion measure (DM) or measured through the radio parallax $\pi$ for the nine pulsars in Table \ref{summ} are summarised in Table \ref{dist}, along with the interstellar extinction along the line of sight.  The values of the 
DM have been assumed from the ATNF pulsar data base and the inferred distance (D$_{\rm DM}$) is based upon the Galactic free electron density model  (Yao et al.\ 2017).  In this case, we assumed an indicative 20\% uncertainty for the distance.
For  PSR\, J1012+5307 (Lazaridis et al.\ 2009) and  PSR\, J1023+0038 (Deller et al.\ 2012) the distance (D$_{\pi}$) is computed from the parallax $\pi$ obtained from radio timing and VLBA interferometry, respectively. In both cases, optical parallaxes  have been obtained from the {\em Gaia} DR2 (Jennings et al.\ 2018), yielding distances of $0.64^{+0.42}_{-0.07}$ kpc and $1.24^{+0.39}_{-0.13}$ kpc, respectively,  which are consistent with those measured from the radio parallaxes. A recent radio timing parallax measurement of PSR\, J1012+5307 gives a distance $D_{\pi} = 1.24\pm 0.24$ kpc (Desvignes et al.\ 2016), similar to that derived by Arzoumanian et al.\ (2018), 
which is within $2\sigma$ from other estimates. 
An optical parallax from {\em Gaia} DR2 has been obtained also for PSR\, B1259$-$63, yielding a distance of $2.21^{+0.19}_{-0.12}$ kpc, which is consistent with that derived from the DM\footnote[7]{See  {\tt http://hosting.astro.cornell.edu/research/parallax/}
for a regularly updated compilation of pulsar parallaxes.}.
For PSR\, J0045$-$7319 we assumed the SMC distance (Graczyk et al.\ 2014) and for PSR\, J0348+0432 the distance based on optical spectroscopy (Antoniadis et al.\ 2013).  For the reddening $E(B-V)$ we assumed the most recent values  reported in the literature.
In the case of PSR\, J0045$-$7319 we assumed the reddening to the SMC (Sabbi et al.\ 2009), whereas for both PSR\, J1810+1744 and PSR\, J2256$-$1024 we adopted the Galactic reddening computed from the Schlegel et al.\ (1998) extinction maps.

\begin{table}
\begin{center}
\caption{Pulsar dispersion measure   (DM), from the ATNF pulsar catalogue (Manchester et al.\ 2005), and distance either inferred from the the DM (D$_{\rm DM}$) and the Galactic free electron density model of Yao et al.\ (2017)
or computed from the radio parallax $\pi$ (D$_{\pi}$), as indicated by the superscript. The fourth column gives the reddening $E(B-V)$ towards  the pulsar as reported in the literature (last column).
}
\label{dist}
\begin{tabular}{lcccc} \hline

Pulsar 				 &  DM & D & $E(B-V)$ & Refs.\\ 
                                           &   (pc cm$^{-3}$) & (kpc)  &  & \\ \hline
J0045$-$7319   	 & 105.4(7) &	 62.1$\pm$1.9$^{\rm *}$  &  0.08$^\dagger$ & 1 \\
B0655+64     		& 8.7738(3) & 0.41$^{\rm DM}$& 0.045 & 2 \\
J1012+5307   	          & 9.02314(5) & 0.82$\pm$0.18$^{\pi}$ & $\approx 0$& 3\\
J1023+0038   	        & 14.325(10) &  1.37$\pm0.04$$^{\pi}$ & 0.1& 4\\
B1259$-$63		& 146.73(1)   &       2.21$^{\rm DM}$            &    0.85 &    5 \\
J1810+1744	         & 39.7  &  2.36$^{\rm DM}$                                    &0.14$^\ddagger$  & 6 \\
J1816+4510   	         & 38.8874(4) & 4.36$^{\rm DM}$ & 0.16--0.32  & 7  \\
J2256$-$1024   	 & 13.8 & 1.33$^{\rm DM}$ & 0.04$^\ddagger$ & 6 \\ \hline
\end{tabular}
\end{center}
1 Sabbi et al.\ (2009); 
2 Kulkarni (1986); 3 Lorimer et al.\ (1995); 4 Patruno et al.\ (2014); 5 Negueruela et al.\ (2011); 6 Breton et al.\ (2013); 7 Kaplan et al.\ (2012) \\
$^{\rm *}$ Distance to the SMC \\ 
 $^\dagger$ Reddening towards the SMC \\ $^\ddagger$ Galactic reddening 
(Schlegel et al.\ 1998)

\end{table}

\section{Conclusions}

Using multi-band photometry measurements from 11 imaging sky surveys we updated the reference to build new SEDs of a sample of nine binary pulsars selected from the ATNF pulsar catalogue.
In particular, for some of them we provided the first possible detections in the near-UV or mid-IR. For five of these pulsars we also inspected the light curves obtained from the Catalina Surveys to look for evidence of flux modulations at the period of the binary system,
only found for PSR\, J1023+0038, whereas for the others the data are either too sparse ot too noisy to look for a statistically significant evidence of such an effect.
The work presented in Paper I and Paper II is set to continue with the the most recent releases of the ATNF pulsar database.

\section*{Acknowledgments}
 Part of this work was written when RPM was hospitalised for a few weeks. His warmest acknowledgments go to Head Nurse Ms. Dorina Terpini of  Nursing Home ``Igea'' (Milan) for securing WI-FI connection for the length of his stay.
We are indebted to Paolo Montegriffo for his support in the use of the GRATIS software.
This research has made use of data obtained from the High Energy Astrophysics Science Archive Research Center  (HEASARC) and the Leicester Database and Archive Service (LEDAS), provided by NASA's Goddard Space Flight Center 
and the Department of Physics and Astronomy, Leicester University, UK, respectively. 
The CSS survey is funded by the National Aeronautics and Space Administration under Grant No. NNG05GF22G issued through the Science Mission Directorate Near-Earth Objects Observations Program.  The CRTS survey is supported by the U.S. National Science Foundation under grants AST-0909182 and AST-1313422.



\begin{thebibliography}{99}


\bibitem[\protect\citeauthoryear{Abdollahi et al.}{2020}]{Abdi2020} Abdollahi S., et al., 2020, ApJS, 247, 33

\bibitem[\protect\citeauthoryear{Abramowski et al.}{2013}]{abra13} H.E.S.S. Collaboration, Abramowski, A., et al., 2013, A\&A, 551, 94 


\bibitem[\protect\citeauthoryear{Antoniadis et al.}{2013}]{anto13} Antoniadis J., et al., 2013, Science, 340, 448 

\bibitem[\protect\citeauthoryear{Archibald et al.}{2009}]{arc09}  Archibald A., et al., 2009, Science, 324, 1411 

\bibitem[\protect\citeauthoryear{Arzoumanian et al.}{2018}]{arz18}Arzoumanian, Z., et al.,  2018, ApJS, 235, 37

\bibitem[\protect\citeauthoryear{Barning}{1963}]{bar63} Barning, F. J. M., 1963, Bull. Astron. Inst. Netherlands, 17, 22

\bibitem[\protect\citeauthoryear{Bell et al.}{1995}]{bell95}  Bell J. F., Bessell M. S., Stappers B. W., Bailes M., Kaspi, V. M.,  1995, ApJ, 447, L117

\bibitem[\protect\citeauthoryear{Bianchi et al.}{2011}]{bia11} Bianchi L.,  Herald  J.,  Efremova B.,  Girardi L.,  Zabot A.,  Marigo P., Conti A.,  Shiao B., 2011, Astrophys. Space Sci., 335, 161 

\bibitem[\protect\citeauthoryear{Bogdanov \& Halpern}{2015}]{bog15}  Bogdanov S. \& Halpern J. P., 2015, ApJL, 803, L27

\bibitem[\protect\citeauthoryear{Bond et al.}{2002}]{bond02} Bond H. E., White R. L., Becker R. H.,  O' Brien M. S., 2002, PASP, 114, 1359 

\bibitem[\protect\citeauthoryear{Bowyer \& Malina}{1991}]{bow91}Bowyer S. \& Malina R. F., 1991, AdSpR, 11, 205 

\bibitem[\protect\citeauthoryear{Breton et al.}{2013}]{bre13} Breton R. P., et al.,  2013, ApJ, 769, 108 


\bibitem[\protect\citeauthoryear{Callanan et al.}{1998}]{cal98}  Callanan P. J., Garnavich P. M., Koester D., 1998, MNRAS, 298, 207 

\bibitem[\protect\citeauthoryear{Christian}{2002}]{chr02}  Christian D. J., 2002, AJ, 124, 3478 

\bibitem[\protect\citeauthoryear{Cordes \& Lazio}{2002}]{2002astro.ph..7156C} Cordes J.~M., \& Lazio T.~J.~W., 2002, arXiv:astro-ph/0207156 

\bibitem[\protect\citeauthoryear{Coti Zelati et al.}{2014}]{cz14} Coti Zelati F., et al., 2014, MNRAS, 444, 1783

\bibitem[\protect\citeauthoryear{Cutri et al.}{2013}]{cutri13}  Cutri R. M., et al., 2013, Explanatory Supplement to the AllWISE Data Release Products

\bibitem[\protect\citeauthoryear{Deller et al.}{2012}]{del12} 	Deller A. T., et al., 2012,  ApJ, 756, L25 
	
\bibitem[\protect\citeauthoryear{Desvignes et al.}{2016}{des16}]	Desvignes, G., Caballero, R. N., Lentati, L., et al.\ 2016, MNRAS, 458, 3341
	
\bibitem[\protect\citeauthoryear{Drake et al.}{2009}]{dra09} Drake A. J., et al.,  2009, ApJ, 696, 870 

\bibitem[\protect\citeauthoryear{Drake et al.}{2014}]{dra14} Drake A. J., et al.,  2014, ApJS, 213, 9 

\bibitem[\protect\citeauthoryear{Fazio et al.}{2004}]{faz04} Fazio G. G., et al., 2004, ApJS, 154, 39 

\bibitem[\protect\citeauthoryear{Fiorentino et al.}{2010}]{fio} Fiorentino G., et al.,  2010, ApJ, 708, 817

\bibitem[\protect\citeauthoryear{Fitzpatrick}{1999}]{fit99} Fitzpatrick E. L. 1999, PASP, 111, 63

\bibitem[\protect\citeauthoryear{Galache et al.}{2008}]{gal08} Galache J. L., Corbet R. H. D., Coe M. J., Laycock S., Schurch M. P. E.,  Markwardt C., Marshall F. E., Lochner J., 2008, ApJS, 177, 189

\bibitem[\protect\citeauthoryear{Jarret et al.}{2011}]{jar11} 	Jarrett T. H., et al., 2011, ApJ, 735, 112 

\bibitem[\protect\citeauthoryear{Jennings et al.}{2018}]{jen18} Jennings R. J., Kaplan D. L., Chatterjee S., Cordes J. M., Deller A. T., 2018, ApJ, 864, 26

\bibitem[\protect\citeauthoryear{Johnston et al.}{1994}]{joh94} Johnston S., Manchester R. N., Lyne A. G., Nicastro L., Spyromilio J., 1994, MNRAS, 268, 430 

\bibitem[\protect\citeauthoryear{Hessels et al.}{2011}]{hes11}  Hessels J. W. T.,  et al. 2011, in Radio Pulsars: An Astrophysical Key to Unlock the Secrets of the Universe, eds. M. Burgay, N. D’Amico, P. Esposito, A. Pellizzoni, \&
A. Possenti,  AIP Conf. Ser. 1357, 40 
 
\bibitem[\protect\citeauthoryear{Hewett et al.}{2006}]{hew06}  	Hewett P. C., Warren S. J., Leggett S. K., Hodgkin S. T., 2006, MNRAS, 367, 454 

\bibitem[\protect\citeauthoryear{Ishihara et al.}{2010}]{ish10} Ishihara D., et al., 2010, A\&A, 514, A1

\bibitem[\protect\citeauthoryear{Kaspi et al.}{1994}]{kas94} Kaspi V. M., Johnston S., Bell J. F., Manchester R. N., Bailes M., Bessell M., Lyne A. G., D'Amico N.,  1994, ApJ, 423, L43 

\bibitem[\protect\citeauthoryear{Kaplan et al.}{2012}]{kap12} Kaplan D. L., et al., 2012, ApJ, 753, 174 

\bibitem[\protect\citeauthoryear{Kaplan et al.}{2013}]{kap13} Kaplan D. L., Bhalerao V. B., van Kerkwijk M. H., Koester D., Kulkarni S. R., Stovall K., 2013, ApJ, 765, 158

\bibitem[\protect\citeauthoryear{Kong}{2013}]{kong13}	Kong A. K. H., 2013, ATel 5515 
	
\bibitem[\protect\citeauthoryear{Kulkarni}{1986}]{kul98} Kulkarni S. R., 1986, ApJ, 306, L85 

\bibitem[\protect\citeauthoryear{Lasker et al.}{2008}]{las08}  Lasker B. M., et al., 2008, AJ,136, 735 

\bibitem[\protect\citeauthoryear{Lawrence et al.}{2007}]{law07}  	Lawrence  A., et al., 	2007, MNRAS, 379, 1599 
 
\bibitem[\protect\citeauthoryear{Lazaridis et al.}{2009}]{laz09}  Lazaridis K., et al, 2009, MNRAS, 400, 805 

\bibitem[\protect\citeauthoryear{Lomb}{1976}]{lom76}  Lomb N. R., 1976, Ap\&SS, 39, 447

\bibitem[\protect\citeauthoryear{Lorimer et al.}{1995}]{lor95} Lorimer D. R., Lyne A. G., Festin  L., Nicastro L., 1995, Nature, 376, 393

\bibitem[Lorimer \& Kramer(2005)]{handbook} Lorimer, D.~R., Kramer, M.\ 2005, {\it Handbook of pulsar astronomy}, Cambridge University Press 

\bibitem[\protect\citeauthoryear{Lynch et al.}{2013}]{lyn13}   Lynch R. S., et al.,  2013, ApJ, 763, 81 
 
\bibitem[\protect\citeauthoryear{Manchester et al.}{2005}]{man05} Manchester R. N., Hobbs G. B., Teoh A. \& Hobbs M., 2005, AJ, 129, 1993 
 
\bibitem[\protect\citeauthoryear{Martin et al.}{2005}]{mar05} Martin D. C., et al., 2005, ApJ, 619, L1 
 
\bibitem[\protect\citeauthoryear{Mason et al.}{2001}]{mas01}  Mason K. O., et al., 2001, A\&A, 365, 36 
 
\bibitem[\protect\citeauthoryear{Mignani et al.}{2014}]{}Mignani R.P., et al.,  2014, MNRAS,  443, 2223 

\bibitem[\protect\citeauthoryear{Miraval Zanon et al.}{2022}]{}Miraval Zanon, A., et al., 2022, A\&A, 660, A63 

\bibitem[\protect\citeauthoryear{Muvareva et al.}{2014}]{muv14} Muraveva T., et al., 2014, MNRAS, 443, 432 
 
\bibitem[\protect\citeauthoryear{Neguerela et al.}{2011}]{neg11} Negueruela I. Rib\'o M., Herrero A.,  Lorenzo J., Khangulyan D., Aharonian F. A., 2011, ApJ, 732, L11 
 
\bibitem[\protect\citeauthoryear{Nicastro et al.}{1995}]{nic95}  Nicastro L., Lyne A. G., Lorimer D. R., Harrison P. A., Bailes M., Skidmore B. D., 1995, MNRAS, 273, L68 

\bibitem[\protect\citeauthoryear{Oke}{1974}]{oke74}  Oke J.B., 1974, ApJS, 27, 21 
 
\bibitem[\protect\citeauthoryear{Onaka et al.}{2007}]{ona07} Onaka T., et al.,  2007, PASJ, 59, 401 
 
\bibitem[\protect\citeauthoryear{Page et al.}{2012}]{page12} Page M. J., et al., 2012, MNRAS, 426, 903 
 
\bibitem[\protect\citeauthoryear{Page et al.}{2014}]{page14} Page M.J., et al., 2014, Proceedings of Swift: 10 Years of Discovery

\bibitem[\protect\citeauthoryear{Papitto et al.}{2013}]{pap13} Papitto, A., et al. 2013, Nature, 501, 517

\bibitem[\protect\citeauthoryear{Papitto et al.}{2019}]{pap19} Papitto A. et al., 2019, ApJ, 882, 104
 
\bibitem[\protect\citeauthoryear{Patruno et al.}{2014}]{pat14}  Patruno A., et al.,  2014, ApJ, 781, L3 
 
\bibitem[\protect\citeauthoryear{Rivera Sandoval et al.}{2018}] {riv18}Rivera Sandoval L. E., et al., 2018, MNRAS, 476, 1086

\bibitem[\protect\citeauthoryear{Roberts}{2013}]{rob13} Roberts M. S. E., 2013, Proceedings of the International Astronomical Union, Volume 291, pp. 127

\bibitem[\protect\citeauthoryear{Roming et al.}{2005}]{rom05} Roming, P. W. A., et al. 2005, Space Sci. Rev., 120, 95 
  
\bibitem[\protect\citeauthoryear{Rajoelimanana et al.}{2011}]{raj11} Rajoelimanana A. F, Charles P.A.,  Udalski A., 2011, MNRAS, 413, 1600
  
\bibitem[\protect\citeauthoryear{Scargle}{1982}]{sca82}  Scargle J. D. 1982, ApJ, 263, 835
  
\bibitem[\protect\citeauthoryear{Schlegel et al.}{1998}]{sch98} Schlegel D. J., Finkbeiner D. P., Davis M., 1998, ApJ, 500, 525

\bibitem[\protect\citeauthoryear{Sollima et al.}{2010}]{sol10} Sollima A., Cacciari C., Bellazzini M., Colucci S., 2010, MNRAS 406, 329

\bibitem[\protect\citeauthoryear{Szkody et al.}{2003}]{szk03} Szkody P.,  et al., 2003, AJ, 126, 1499 

\bibitem[\protect\citeauthoryear{Skrutskie et al.}{2006}]{skr06}  Skrutskie M.~F., et al.\ 2006, AJ, 131, 1163  

\bibitem[\protect\citeauthoryear{Stappers et al.}{2014}]{sta14}  Stappers B., 2014, ApJ, 790, 39 
  
\bibitem[\protect\citeauthoryear{Stair et al.}{2001}]{sta01}   Stairs I. H., et al., 2001, MNRAS, 325, 979 
  
\bibitem[\protect\citeauthoryear{Shklovskii}{1970}]{shk70}  Shklovskii, I. S. 1970, Soviet Ast., 13, 562

\bibitem[\protect\citeauthoryear{Stephenson C. B. \& Sanduleak}{1971}]{ste71}  Stephenson C. B. \& Sanduleak N., 1971, Publication of the Warner and Swasey Observatory, Cleveland, Ohio: Case Western Reserve University 

\bibitem[\protect\citeauthoryear{Yershov}{2014}]{yershov14} Yershov, V.N., 2014, Astrophys. \& Space Sci., 394, 57
 
\bibitem[\protect\citeauthoryear{York et al.}{2000}]{york00}   York D.G., et al.,  2000, AJ, 120, 1579 
  
\bibitem[\protect\citeauthoryear{Takata et al.}{2014}]{tak14}   Takata J., et al., 2014, ApJ,  785, 131 
  
\bibitem[\protect\citeauthoryear{Tendulkar et al.}{2014}]{ten14}  Tendulkar S. P., et al., 2014, ApJ, 791, 77   

\bibitem[\protect\citeauthoryear{Teplitz et al.}{2010}]{tep10} Teplits H. I. et al.\ 2010, Astronomical Data Analysis Software and Systems XIX.  ASP Conference Series, Vol. 434, p.437
  
\bibitem[\protect\citeauthoryear{Thorstensen \& Armstrong}{2005}]{tho05}   	Thorstensen J. R. \& Armstrong E.,  2005, AJ, 130, 759 
	
\bibitem[\protect\citeauthoryear{van Kerkwijk \& Kulkarni}{1995}]{vdkk95}    van Kerkwijk M. H. \& Kulkarni S. R., 1995, ApJ, 454, L141 
  
\bibitem[\protect\citeauthoryear{van Kerkwijk et al.}{1996}]{vdk96}    van Kerkwijk M. H., Bergeron P., Kulkarni S. R., 1996, ApJ, 467, L89 
  
\bibitem[\protect\citeauthoryear{Yao et al.}{2017}]{yao17} Yao J. N., Manchester R. N., Wang N., 2017, ApJ, 835, 29
  
\bibitem[\protect\citeauthoryear{Wang et al.}{2013}]{wang13}    Wang X.,  Wang Z., Morrell N., 2013, ApJ, 764, 144

\bibitem[\protect\citeauthoryear{Webb et al.}{2004}]{webb04} Webb N. A., Olive J.-F., Barret D., Kramer M., Cognard I., L\"ohmer O., 2004, A\&A, 419, 269
  
\bibitem[\protect\citeauthoryear{Werner et al.}{2004}]{wer04}    Werner M. W., Gallagher D. B.,  Irace W. R., 2004, AdSpR, 34, 600 
  
\bibitem[\protect\citeauthoryear{Woudt et al.}{2004}]{wou04}   Woudt P. A., Warner B., Pretorius M. L., 2004, MNRAS, 351, 1015 

\bibitem[\protect\citeauthoryear{Wright et al.}{2010}]{wri10}   Wright E. L., et al.,  2010, AJ, 140, 1868 
    
\end{thebibliography}

\label{lastpage}

\end{document}